\documentclass[13pt]{article}
\usepackage{amssymb, amsmath, amsbsy} 
\usepackage{slashed}
\usepackage[filecolor=Gri,urlcolor=azc]{hyperref}
\hypersetup{colorlinks=false}
\hypersetup{pdfborder={0 0 0}}
\usepackage{pdfpages}
\usepackage{epstopdf}
\usepackage{float}
\usepackage[subfigure]{graphfig}
\usepackage[affil-it]{authblk}

\usepackage[left=2cm,right=2cm,top=3cm,bottom=3cm]{geometry}
\usepackage{xcolor}
\begin{document} 

\author[1]{M. A. Arroyo-Ure\~na \thanks{\texttt{marcofis@yahoo.com.mx}}}
\author[2,3]{J. Lorenzo D\'iaz-Cruz\thanks{\texttt{jldiaz@fcfm.buap.mx}}}
\author[2,4]{Bryan O. Larios-L\'opez\thanks{\texttt{bryanlarios@gmail.com}}}
\author[3]{M. A. P\'erez de Le\'on\thanks{\texttt{marioaldair\_07@hotmail.com}}}
\title{\textbf{A Private SUSY 4HDM with FCNC in the Up-sector}}
\affil[1]{Facultad de Estudios Superiores Cuautitlan -UNAM 
\protect\\ Cuautitlan, Edo de Mex.,  M\'exico} 
\affil[2]{Mesoamerican Center for Theoretical Physics, UNACH, Chiapas, M\'exico } 
\affil[3]{CIFFU and Facultad de Ciencias F\'isico - Matem\'aticas, 
\protect\\ Benem\'erita Universidad Aut\'onoma de Puebla 
\protect\\Apdo. Postal 1364, C.P. 72000, Puebla, Pue. M\'exico} 
\affil[4]{Departamento de Gravitaci\'on y Altas Energ\'ias, Facultad de Ciencias
\protect\\Universidad Nacional Aut\'onoma de Honduras
\protect\\Ciudad Universitaria, Tegucigalpa M.D.C. Honduras}
\maketitle
\begin{abstract}
	We present a SUSY  model with four Higgs doublets of the ``private type", where each fermion
	type (up, down, and charged leptons) obtain their masses from a different Higgs doublet
	$H_f\,( f = u_1, d, e)$. The conditions for  anomaly cancellation, imply that the remaining Higgs doublet of the model  ($H_{u_2}$), must have the same hypercharge as $H_{u_1}$, and thus can only couple to up-type quarks, which opens the possibility to have FCNC's only  in this sector.  We study the Lagrangian  of the model, and in particular the Higgs potential, in order to identify the Higgs mass-eigenstates and their interactions; for the Yukawa matrices we consider the 4-texture case. We obtain constraints on the model parameters by using LHC measurements on  the properties of the 125 GeV Higgs boson ($h$),  and identify viable regions of parameter space. These constraints are then used to evaluate the prospects to detect the FCNC decay mode $t \to ch$ at the future high-luminosity (HL) option for the LHC, which are compared with current limits from LHC-run2. Then, we also evaluate the FCNC decay of the next heavier Higgs boson  $H_2\to tc$, 
which  can reach typically  $BR(H_2  \rightarrow tc) \approx \mathcal{O}(10^{-4}-10^{-5} )$; the search for the signal at HL-LHC is also studied, 
finding that it could be detectable for some specific regions of the  model parameter space.
\end{abstract}   

\unitlength=1mm

\newpage
\section{Introduction}\label{introduction}
After the Higgs discovery at the LHC \cite{lhc1,lhc2,lhc3}, the attention has focused on the precision test of the Higgs boson properties. So far, the measurements of its couplings with fermions and gauge bosons point towards a SM interpretation, however given that only a few of them have been probed \cite{Delaunay:2013pja}, there is still room for new physics. In particular, regarding the fermion-Higgs couplings, LHC has measured directly or indirectly, only the coupling with top, bottom and tau pairs, and all of them seem to fit the SM predictions. Namely, these couplings lay on a straight line, as function of the fermion mass \cite{Arroyo-Urena:2020fkt}. However, the precision level achieved so far allows for other possibilities, for instance it could be that each fermion type  acquires its mass from  its own ``private'' Higgs \cite{theprivate,privatelhc,DCUS}, and then the corresponding Higgs-fermion couplings would lay on different lines.  Extension of the Higgs sector not only predict modification of the SM couplings, but could also include new type of interactions, such as FCNC Higgs fermions couplings, as well as rich spectrum of heavier neutral and charged Higgs particles.

In particular, multi-Doublet Higgs Models are a straightforward extension of the SM, where the total vacuum expectation value (vev) is given by $v^{2}=v_{1}^{2}+v_{2}^{2}+\cdots +v_{n}^{2}=(246\:  \mathrm{GeV})^{2}$, and $v_{i}=\langle  0 \lvert H^0_{i} \lvert 0\rangle$ is the vev of the neutral component of each Higgs doublet $H_i$ ($i=1,2,..,n$) \cite{Branco:2011iw}. In this case there are deviations from the SM predictions for the Higgs couplings $hWW$, $hZZ$, with $h=h_{1}^{0}$  being the SM-like Higgs boson, i.e. the lightest neutral $\mathrm{CP}-$even Higgs within the scalar spectrum.
This type of models offer also the possibility of having flavor-changing neutral Higgs couplings, which have been studied in the past, including the case where fermion hierarchy is reproduced through the FN mechanism \cite{Tsumura:2009yf,DCBG}. 
Within the most general version of the 2HDM, both Higgs doublets couple to all types of fermions. In this case, the diagonalization of the full mass matrix, does not imply that each Yukawa matrix is diagonalized with the same rotations, therefore FCNC can appear at tree level. Within this general model, one must reproduce the observed fermion masses and mixing angles, while at the same time the level of FCNC must satisfy current experimental bounds \cite{fcnc2}. One possibility to achieve this, is the assumption that the Yukawa matrices have a certain texture form, i.e. with zeros in different elements, and in particular it is known that the 4-zero texture is consistent with data from flavor physics.

Supersymmetry (SUSY) has also been widely studied as a possibility to solve or at least to ameliorate, the hierarchy 
problem \cite{Martin:1997ns}. The minimal SUSY model (MSSM) includes two Higgs doublets, and its structure is such that each doublet couples to only one type of fermion, and thus FCNC are not allowed in the model. The next multi-doublet SUSY Higgs model must include four Higgs doublets, where each doublet is denoted as $H_i (i=1,2,3,4)$ \cite{drees,masip}; in this case we can have more possibilities for flavor physics \cite{Aranda:2000zf,DiazCruz:2002er}. This type of model is also motivated, for instance, from considerations from $LR$ symmetry or unification \cite{Dutta:2018yos}. The phenomenology of this model has been studied recently in refs.  \cite{kawase,kanemura,Gupta:2009wn}, which depends on a large number of free-parameters. \\

In this paper we shall consider a four-Higgs doublet SUSY model model with a more restricted parameter space; this is achieved by considering a version of the model of the "private type". This model is defined  by requiring that one doublet gives masses to each fermion type, namely, $H_{1}\equiv H_{u_1}$ gives mass to up-type quarks, $H_{2}\equiv H_d$ gives masses to down-type quarks and $H_{3}\equiv H_l$ gives mass to charged leptons. Then we have an extra  Higgs doublet left ($H_4$), however this doublet  should have the same hypercharge as $H_{u_1}$, in order for the anomalies to be canceled.  Therefore it could only couple to up-type quarks, and we call this doublet as $H_4\equiv H_{u_2}$. Thus, in this ``private" SUSY Higgs model, we could only have FCNC in the up-type quarks, which would predict that the decays $t\rightarrow ch$ and $H^0_i\rightarrow t \overline{c}$ occur at some level, where $h=H^0_1$ and $H^0_i$ are part of the Higgs spectrum.

The goal of this paper is to construct this private SUSY Higgs model,  and derive the interactions of Higgs boson with gauge bosons and fermions. Then, we want to identify regions of parameter space that are consistent with high energy data on the Higgs couplings, as derived from LHC. As in our model, FCNC only occur in the up-type quark sector, one would need to consider  the constraints from low-energy flavor physics, however they are rather mild. In fact, the LHC data on Higgs couplings to gauge bosons and fermions (including FCNC top decay $t\rightarrow ch$) provide stronger constraints \cite{AguilarSaavedra:2004wm}.
These constraints are then used to evaluate the FCNC decay of the next heavier Higgs boson  $H\rightarrow t \bar{c}$. 

The organization of this paper goes as follows. In Section 2 we review the construction of the model, including the Higgs potential (for this we follow closely Ref. \cite{kawase}), then we perform the minimization of the Higgs potential ($V$) and derive the scalar mass matrices; we also present an approximate diagonalization of the $4\times4$ matrix, including the leading radiative corrections to the Higgs mass coming from the 
top-stop system, using the effective potential technique. We also discuss the Yukawa Lagrangian, assuming a four-texture Yukawa matrix, and derive the interactions of the Higgs bosons with the fermions and the gauge boson. Section 3 contains the analysis of Higgs couplings and the constraints obtained from LHC. Section 4 includes our study of FCNC decay for the lightest Higgs boson $t\rightarrow ch$, as well as the FCNC decay  of the next heavier neutral Higgs boson $H_2  \rightarrow t \bar{c} $, including a signal-vs-background study to determine  the detectability of the signals at  LHC.  Finally our conclusions are presented in the section 5.

\section{The  SUSY four-Higgs doublet model}

The minimal supersymmetric extension of the Standard Model (MSSM) includes two Higgs doublets, and it can be extended 
by enlarging its Higgs content. In particular, we are interested in studying a supersymmetric version of the 
four Higgs doublet model (SS-4HDM)  \cite{kawase}. The gauge symmetry is the same as in the SM, thus the model
includes the usual MSSM gauge bosons and gauginos. Similarly, the model includes the
same  MSSM particle content for quarks, leptons, and their superpartners (squarks and sleptons). 
Thus, the fermion-sfermion interactions will be almost the same as in the MSSM, and moreover the possible modifications 
coming from a modified Yukawa sector, shall be neglected in a consistent manner, as we shall explain next.

\subsection{The Superpotential of the Model} 

The starting point when studying the Higgs and Yukawa sectors in a Supersymmetric model, is to describe 
the corresponding superpotential. This model includes four Higgs chiral  Superfields, each of them  includes both the scalar 
and the fermion components, namely the Higgses and the Higgsinos. Then,  because of anomaly cancellation,
two of them, $\hat{H}_{u1}, \hat{H}_{u2}$, should have weak hypercharge $Y=-1$ ,  while the other two, $\hat{H}_{d}$ 
and $\hat{H}_l$ must have $Y=+1$. The general superpotential includes the usual MSSM Yukawa term ($W_{Yuk}$) and the mu-terms, 
which now  takes the form: 

\begin{equation}
W= W_{Yuk} + \sum_i [ \mu_ {i1}  \hat{H}_{ui} \cdot \hat{H}_{d} + \mu_ {i2}  \hat{H}_{ui} \cdot \hat{H}_{l}] 
\end{equation}


Then, one can follow the rules of supersymmetry to derive the scalar Higgs potential. Namely, one derives from the 
superpotential the corresponding F-terms, then adds the D-terms. Finally, one needs to include the allowed SUSY 
soft-breaking terms.

In turn, the form of the Yukawa superpotential ($W_{Yuk}$) depends on whether one accepts or 
not some level of flavor-changing  neutral currents (FCNC) in the model, which should come with a suppression mechanism 
in order to satisfy current FCNC bounds.  However, as we are interested in the class of models where the mass of the leptons, up- and down-quarks 
come from their private Higgs doublet, the possibilities are quite limited.  Furthermore, given the assignment of hypercharges, 
it turns out 
that the model is quite constrained. Namely, if the mases of down-quarks and leptons come from two of the Higgs doublets, namely
$\hat{H}_{d}$ and $\hat{H}_l$ 
respectively, then the only possibility left is to have FCNC in the up-quark sector, because the
remaining Higgs superfields  $\hat{H}_{u1}, \hat{H}_{u2}$ are the ones that could couple with the up-type quarks 
(and up-type squarks).
Thus, the Yukawa superpotential of our model is of the form:

\begin{equation}
W_{Yuk} =  [ \hat{Q} Y_{u1}  \hat{H}_{u1} \hat{U} + \hat{Q} Y_{u2}  \hat{H}_{u2} \hat{U}]  + \hat{Q} Y_ {d}  \hat{H}_d  \hat{D} +  \hat{L} Y_l \hat{H}_{l} \hat{E}  
\end{equation}

There are many possibilities to motivate the private Higgs assignments, for instance one can use some set of discrete symmetries.   Thus the mass matrix for up-type quarks will receive contributions from two Yukawa matrices, which will induce FCNC Higgs couplings for up-type quarks  \footnote{In the original ``Private Higgs model", not only each fermion type gets its mass from different Higgs doublet, but also the Yukawa couplings are of $\mathcal{O}(1)$, but then to achieve a realistic spectrum  one needs to introduce extra fields and the Froggatt-Nielsen mechanism. Here we choose to leave this aside, and just concentrate on the phenomenological implications of the simplest model.}. 
These assignments can be obtained in many ways, for instance we can incorporate a set of discrete symmetries ($k_u, k_d, k_l$) with parities shown in Table 1.

\begin{table}[h!]
	\begin{center}
		\label{tab:table1}
		\begin{tabular}{c|c|c|c|c|c}
			
			&$H_u,\bar{u}$ & $H_d,\bar{d}$& $H_l,\bar{l}$& $Q$ &$L$\\
			\hline
			$k_u$ & $-$ & + & + & +& +\\
			$k_d$ & + & $-$ & + & +& +\\
			$k_l$ & $-$ & + & $-$ & +& +\\
			\hline
		\end{tabular}
		\caption{Discrete symmetries for our SUSY 4HDM.}
	\end{center}
\end{table}

The power of these discrete symmetries is such that it permits to classify the different types of multi-Higgs models. For instance, when one talks about a Two-Higgs doublet model (2HDM), one can discuss the types I, II or III; in this model both doublets $H_u$, $H_d$ have the same quantum numbers, so in principle both of them could couple to all types of fermions, and type III would be mandatory. But thanks to the choice of appropriate discrete symmetries, one can build the model where only one Higgs doublet couple to each fermion type (2HDM of type II) or where only one doublet does the job of generating all fermion masses (as in the 2HDM  type I). Similarly, we can use the discrete symmetries of Table 1 to define our ``SUSY Private Higgs model", which only allows FCNC in the Up sector, and not for the down type quarks and leptons.  Other discrete symmetries could be use to forbid completely FCNC, but having a fourth Higgs doublet with the right quantum numbers make more natural  to explore possible FCNC in the Up type quarks.

It is possible that the discrete symmetries could make some of the mu-terms to vanish. Furthermore, these discrete symmetries could also be imposed on the  soft-breaking terms. Then, the question surges on whether the resulting Higgs
spectrum is viable, namely that one can have a light SM-like Higgs boson with $m_h=125$ GeV, accompanied with heavier neutral and
charged  Higgs bosons, heavy-enough to satisfy current bounds from LHC on extra Higgs particles.  This will be done in the next section, after discussing the minimization of the Higgs potential and the Higgs mass matrices.

\subsection{The Higgs potential of the SS-4HDM}

In this section we shall discuss the general Higgs potential for the supersymmetric model with four Higgs Doublets, following the notation
of Ref. \cite{kawase}. After discussing the minimization conditions, the general form of the Higgs mass matrices will be derived. 

The Higgs doublets are the scalar components of the Higgs chiral multiplets, and are written as follows:
\begin{equation}
H_{ui} = 
\begin{pmatrix}
H_{ui}^+\\
H_{ui}^0
\end{pmatrix}
(i=1,2),\hspace{0.5cm}
H_{d} = 
\begin{pmatrix}
H_{d}^0\\
H_{d}^-
\end{pmatrix},
\hspace{0.5cm}
H_{l} = 
\begin{pmatrix}
H_{l}^0\\
H_{l}^-
\end{pmatrix},
\end{equation}
where $H^0_k$, is given by
\begin{equation}
H^0_k=\frac{1}{\sqrt{2}}(v_k+\eta_k+i\chi_k)\hspace{0.5cm}\mathrm{and}\hspace{0.5cm} (k=u_1,d,l,u_2),
\end{equation}\label{eq:Higgs_field_1}

As in the MSSM, $H_{ui}$ should have an hypercharge equal to  -1, while for $H_{d,l}$ it is +1. The Higgs doublets with hypercharge +1 give mass to the down type quarks sector $(H_d)$ and the charged leptons $(H_l)$. \\

As it was show in \cite{kawase,kanemura}, the scalar potential of the Higgs fields takes the following form
\begin{align}\label{eq:Higgs_potential}\nonumber
V&=\sum_{i=1}^2\bigg((|\mu_{i1}|^2+|\mu_{i2}|^2+\tilde{m}_{ui}^2)(|H_{ui}^0|^2+|H_{ui}^+|^2)+(|\mu_{1i}|^2+|\mu_{2i}|^2+\tilde{m}_{di}^2)(|H_{di}^0|^2+|H_{di}^+|^2)\bigg)\\
&\quad + \bigg((\mu_{11}^*\mu_{21}+\mu_{12}^*\mu_{22})(H_{u1}^{0*}H_{u2}^{0}+H_{u1}^{+*}H_{u2}^{+})+(\mu_{11}^{*}\mu_{12}+\mu_{21}^{*}\mu_{22})(H_{d1}^{0*}H_{d2}^{0}+H_{d1}^{-*}H_{d2}^{-})+c.c\bigg)\nonumber\\
&\quad+\Bigg(\sum_{i=1}^{2}\sum_{j=1}^{2}b^2_{ij}(H_{ui}^{+}H_{dj}^{-}-H_{ui}^{0}H_{dj}^{0})+c.c.\Bigg)+\frac{g^2+g'^2}{8}\Bigg(\sum_{i=1}^{2}(|H_{ui}^{0}|^2+|H_{ui}^{+}|^2-|H_{di}^{0}|^2+|H_{di}^{-}|^2)\Bigg)^2\nonumber\\
&\quad+\frac{g^2}{2}\bigg(\sum_{i=1}^{2}|(H_{ui}^{+*}H_{ui}^{0}+H_{di}^{0*}H_{di}^{-})|^2-\sum_{i=1}^{2}\sum_{j=1}^{2}(|H_{ui}^{0}|^2-|H_{di}^{0}|^2)(|H_{uj}^{+}|^2-|H_{dj}^{-}|^2)\bigg),
\end{align}
here $H_{d_2}=H_l$. In general the parameters $\mu_{ij}$, $\tilde{m}_{ui,di}$ and $b_{ij}$ could be complex, but for simplicity throughout this paper we shall take them to be  real parameters. The vev's are parametrized in terms of the total SM vev ($v$) and the angles $\alpha,\beta$ and $\omega$ as follows:
\begin{align}\label{eq:vev_2}
v_{1}&= \frac{\sqrt{2}M_Z}{\sqrt{(g^2+g'^2)(1+\tan^2\omega)}}\cos\beta,\\
v_{4}&= \frac{\sqrt{2}M_Z}{\sqrt{(g^2+g'^2)(1+\tan^2\omega)}}\tan\omega\sin\alpha,\\
v_d&= \frac{\sqrt{2}M_Z}{\sqrt{(g^2+g'^2)(1+\tan^2\omega)}}\sin\beta,\\
v_l&= \frac{\sqrt{2}M_Z}{\sqrt{(g^2+g'^2)(1+\tan^2\omega)}}\tan\omega\cos\alpha.
\end{align}
The minimization conditions $\frac{\partial V}{\partial H_{ui}^0}=0$ and  $\frac{\partial V}{\partial H_{di}^0}=0$ evaluated in the VEVs take the following form: 
\begin{align}\label{eq:minimization_condition}
\Delta_{u_1}+\left(\mu _{11} \mu _{21}+\mu _{12} \mu _{22}\right)c^{-1}_\beta s_\alpha t_\omega - b^2_{12}c^{-1}_\beta c_\alpha t_\omega- b^2_{11} t _\beta+\frac{1}{4}\text{Mz}^2\left(c_{2 \beta}c^2_\omega
-c_ {2 \alpha}s ^2_\omega\right)&=0,\\
\Delta_d+\left(\mu _{11} \mu _{12}+\mu _{21}
\mu _{22}\right)s^{-1}_\beta c_\alpha t_\omega-b^2_{21}s^{-1}_\beta s_\alpha t_\omega- b^2_{11}t^{-1}_\beta+\frac{1}{4}\text{Mz}^2 \left( c_{2 \alpha} s ^2_\omega -c_{2\beta}c ^2_\omega\right)&=0,\\
\Delta_l - b^2_{21}s_\beta t^{-1}_\omega s^{-1}_\alpha+ \left(\mu _{11} \mu _{21}+\mu _{12} \mu _{22}\right)c_\beta t^{-1}_\omega s^{-1}_\alpha-b^2_{22}t^{-1}_ \omega s^{-1}_\alpha c_\alpha+\frac{1}{4}\text{Mz}^2 s_\omega\left[c_{2\beta} c_\omega-c_{2\alpha} s_\omega\right]&=0,\\
\Delta_{u_2}  - b^2_{12}c_\beta c^{-1}_\alpha t^{-1}_\omega-(\mu_{11}\mu_{12}+\mu_{21}\mu_{22})s_\beta c^{-1}_\alpha t^{-1}_\omega-t_\alpha t^{-1}_\omega b^2_{22}+\frac{1}{4}\text{Mz}^2\left[c_{2\alpha}s^2_\omega-c_{2\beta}c^2_\omega\right]&=0,
\end{align}
where $\Delta_{u1}=\mu_{11}^2+\mu_{12}^2+\tilde{m}_{u1}^2$, $\Delta_{d}=\mu_{11}^2+\mu_{21}^2+\tilde{m}_{d}^2$, $\Delta_{u2}=\mu_{21}^2+\mu_{22}^2+\tilde{m}_{u2}^2$, $\Delta_{l}=\mu_{12}^2+\mu_{22}^2+\tilde{m}_{l}^2$.  
Here we have renamed $\tilde{m}_{d2} = \tilde{m}_{l}$. This will be used next in the construction of the Higgs mass matrices.

\subsection{Higgs Mass Matrices}
We shall consider the CP-invariant case, and focus on  the neutral CP-even Higgs fields, which are obtained from
the real parts of the neutral components. In the basis ($H_{u1}^0,H_{d1}^0, H_{l}^0, H_{u2}^0,$), 
the mass matrix can be written as follows
\begin{equation}\label{eq:Higgs_mass_real}
\mathcal{M}^2 = 
\begin{pmatrix}
m_{u_1u_1}^2 & m_{u_1d}^2 & m_{u_1l}^2 & m_{u_1 u_2}^2\\
m_{du_1}^2 & m_{dd}^2 & m_{dl}^2 & m_{u_2 d}^2\\
m_{lu_1}^2 & m_{ld}^2 & m_{ll}^2 & m_{u_2 l}^2\\
m_{u_2 u_1}^2 & m_{u_2 d}^2 & m_{u_2 l}^2 & m_{u_2u_2}^2
\end{pmatrix},
\end{equation}
where each element of the matrix (Eq.~(\ref{eq:Higgs_mass_real})) takes the form
\begin{align}
m^2_{u_1u_1}&=\hspace{0.3cm}\frac{1}{2}\Delta_{u1}+\frac{1}{8}m_Z^2\left((2 \cos2\beta+1)\cos ^2\omega-\cos2\alpha\sin ^2\omega 
\right),\\
m^2_{u1d}&=-\frac{1}{2} b^2_{11}-\frac{1}{2}m_Z^2\sin 2\beta \cos ^2\omega,\\
m^2_{u_1l}&=\hspace{0.3cm}\frac{1}{2}\left(\mu_{11} \mu _{21}+\mu _{12}\mu _{22}\right)+\frac{1}{2}m_Z^2\cos\beta\sin\alpha \sin 2\omega,  \\
m^2_{dd} &=\hspace{0.3cm}\frac{1}{2}\Delta_d+\frac{1}{8}m_Z^2 \left((1-2 \cos2 \beta) \cos ^2\omega+\cos 2 \alpha \sin
^2\omega\right),\\
m^2_{dl}&=-\frac{1}{2}b^2_{21}-\frac{1}{2} m_Z^2\sin\alpha\sin\beta\sin 2\omega,\\
m^2_{ll}&=\hspace{0.3cm}\frac{1}{2}\Delta_l+ \frac{1}{8}m_Z^2\left((1-2 \cos 2\alpha) \sin ^2\omega+\cos 2\beta\cos ^2\omega\right),\\
m_{u_2u_2}^2&=\hspace{0.3cm}\frac{1}{2}\Delta_{u_2}+\frac{1}{8}m_Z^2\left((2\cos{2\alpha}+1)\sin^2\omega-\cos{2\beta} \cos^2\omega\right),\\
m_{u_2u_1}^2&=-\frac{1}{2}b^2_{12}-\frac{1}{2}m_Z^2\cos\alpha\cos\beta\sin 2\omega,\\
m_{u_2 d}^2&=\hspace{0.3cm}
\frac{1}{2}(\mu_{11}\mu_{12}+\mu_{21}\mu_{22})+\frac{1}{2}m_Z^2\cos\alpha\sin\beta\sin 2\omega,\\
m_{u_2 l}^2 & = -\frac{1}{2}b^2_{22}-\frac{1}{2}m_Z^2\sin 2\alpha \sin^2\omega.
\end{align}
Since the CP-even Higgs mass matrix is symmetric, we only need to give the ten independent  components.

\bigskip


\subsection{Approximate Diagonalization}

Here we shall assume that the $i4$ ($i=1,2,3$) entries of the mass matrix are small as compared with the remaining entries, namely
\begin{equation}
\mathcal{M}^2 \approx 
\begin{pmatrix}
m_{u_1u_1}^2 & m_{u_1d}^2 & m_{u_1l}^2 & \epsilon_1\\
m_{du_1}^2 & m_{dd}^2 & m_{dl}^2 & \epsilon_2\\
m_{lu_1}^2 & m_{ld}^2 & m_{ll}^2 & \epsilon_3\\
\epsilon_1 & \epsilon_2 & \epsilon_3 & m_{u_2u_2}^2
\end{pmatrix}
\end{equation}
Thus we can perform an approximate diagonalization, i.e. the matrix $O(\delta_i)$ is defined as follows.
\begin{equation}\label{eq:Diag}
\begin{pmatrix}
h_1\\
h_2\\
h_3\\
h_4\\
\end{pmatrix}
=
O(\delta_i)
\begin{pmatrix}
\eta_1\\
\eta_d\\
\eta_l\\
\eta_2	\\
\end{pmatrix},
\end{equation}
and
\begin{equation}
\mathcal{M}^2=O^T(\delta_i)
\begin{pmatrix}
m^2_{h_1}&0&0&0\\
0&m^2_{h_2}&0&0\\
0&0&m^2_{h_3}&0\\
0&0&0&m^2_{h_4}\\
\end{pmatrix}
O(\delta_i).
\end{equation}
The $O(\delta_i)$ matrix will have the following form (with $i=1,2,3$) \cite{diago}
\begin{equation}
O^{T}(\delta_i)=
\begin{pmatrix}
c_{\delta _1}c_{\delta _2}& -c_{\delta _3
}s_{\delta _1}-c_{\delta _1}s_{\delta _2}s_{\delta _3}& s_{\delta _1}s_{\delta _3}-c_{\delta _1}c_{\delta _3} s_{\delta _2} & 0 \\
c_{\delta _2}s_{\delta _1} & c_{\delta _1}c_{\delta _3}-s_{\delta _1}s_{\delta _2}s_{\delta _3}& -c_{\delta _3}s_{\delta _1}s_{\delta _2}-c_{\delta _1}s_{\delta _3} & 0 \\
s_{\delta _2} & c_{\delta _2}s_{\delta _3} & c_{\delta _2}c_{\delta _3}& 0\\
0 & 0 & 0 & 1\\
\end{pmatrix}\label{matrixO}
\end{equation}

Then, we obtain the following expressions for $\eta_i$ fields, written in terms of the mass eigenstates $H_i$ ($i=1,4$):

\begin{align}
\eta_1&=c_{\delta _1}c_{\delta _2}h_1 -(c_{\delta _3
}s_{\delta _1}+c_{\delta _1}s_{\delta _2}s_{\delta _3} )h_2+ (s_{\delta _1}s_{\delta _3}-c_{\delta _1}c_{\delta _3} s_{\delta _2} )h_3,  \\
\eta_d&=c_{\delta _2}s_{\delta _1}h_1 +(c_{\delta _1}c_{\delta _3}-s_{\delta _1}s_{\delta _2}s_{\delta _3})h_2 -(c_{\delta _3}s_{\delta _1}s_{\delta _2}+c_{\delta _1}s_{\delta _3})h_3,\\
\eta_l&=s_{\delta _2}h_1 +c_{\delta _2}s_{\delta _3}h_2 +c_{\delta _2}c_{\delta _3}h_3,\\
\eta_4&=h_4.
\end{align}

These expressions will be used next to derive the Higgs interactions with fermions and gauge bosons.

\subsection{The Higgs boson spectrum}

Now, we should verify that  the resulting Higgs spectrum is viable, 
namely that we have a light SM-like Higgs boson with $m_h=125$ GeV, accompanied with heavier neutral and
charged  Higgs boson, in agreement with current LHC bounds on Higgs particles. 
In order to explore the model parameters and symmetries, we shall consider the case when the discrete symmetries
for the Yukawa sector, also apply for the supersymmetric $\mu$-terms, i.e. $\mu_{ij}=0$.  Here it will be relevant to include the
leading SUSY radiative corrections to the Higgs mass, which are dominated by the top-stop loops. These radiative corrections depend on
the stop-top interactions, and for them we shall neglect the flavor-violating interactions, as we know that the current bounds on FCNC
processes makes them to be small, which is a consistent approximation.  Following the effective potential technique to include the leading radiative stop-top corrections, we obtain  the following expression for the mass of the lightest Higgs boson ($h (=H_1)$):

\begin{eqnarray}
m^2_{h}&=& \frac{m_Z^2}{8} F(\alpha, \beta, \delta_i, \omega)  
-  s_{\delta_1} c_ {\delta_1} c^2_{\delta_2} b^2_{11} -   s_{\delta_1} s_{\delta_2} c_{\delta_2} b^2_{21} \\
&&+\frac{1}{2} \tilde{m}_{u_1}^2 c^2_{\delta_1} 
c^2_{\delta_2}+\frac{1}{2} \tilde{m}_d^2 s^2_{\delta_1} c^2_{\delta_2}+\frac{1}{2} \tilde{m}_l^2 s^2_{\delta_2}
+\frac{3G_F}{\sqrt{2}\pi^2}\frac{m_t^4}{r_{1u}}\log\frac{m^2_{stop}}{m^2_t}\nonumber
\end{eqnarray}

where $r_{1u}= v^2_1/v^2$ and:

\begin{eqnarray}
F (\alpha, \beta, \delta_i, \omega) & =&  \left[ \right. c^2_{\delta_1} c ^2_{\delta_2} \left( (2 c_{ 2 \beta} +1) c^2_{\omega}  - c_{ 2 \alpha}  s^2_\omega \right)
+ s ^2_{\delta_1} c ^2_{\delta_2} \left( c_{ 2 \alpha}  s ^2_\omega +(1-2 c_{ 2 \beta} ) c ^2_\omega\right)  
+s ^2_{\delta_2} \left( (1-2 c_{ 2 \alpha} ) s ^2_\omega +c_{ 2 \beta}  c^2_\omega \right) 
\nonumber \\
& &+c_ \beta  c_{\delta_1} \left( s_{\alpha}  s_{ 2 \delta_2} s_{ 2 \omega} 
- 4 s_ {\beta}  s_{\delta_1} c ^2_{\delta_2}c ^2_\omega \right) 
-4 s_\alpha  s_\beta  s_{\delta_1} s_{\delta_2} c_{\delta_2} s_\omega  c_\omega \left.\right] 
\end{eqnarray}

Similar formulas are derived for the mases of the heavier CP-even Higgs bosons.

Then, in Fig. 1a we show  $m_h$ as a function of  $\tilde{m}_{d}$, for the stop mass: 
$m_{stop} = 1\hspace{0.1cm}$ TeV, the different lines correspond to the soft Higgs mass parameter: 
$\tilde{m}_{u1}=300, 400, 500, 600, 700$ GeV,   and for the  fixed value $\tilde{m}_l =500$ GeV, while
the  b-terms are fixed as: $b_{11}=550$ GeV, $b_{21}=50GeV$. The relevant angles are chosen as:
$\alpha=0.01, \, \beta=0.45, \, \omega=0.37, \, \delta_1=0.31, \, \delta_2=0,37, \delta_3 = 2.1$. 
\footnote{These values of the angles are actually chosen such that the couplings of the light SM-like
Higgs boson $h$, resemble the SM values, as it was actually observed by LHC; this will be discussed in detail in Section 4.}
We can see that  the experimental value $m_{h} \simeq 125$ GeV (solid horizontal line) crosses with all the lines shown 
in figure 1a.


Next we show in figures 1b-c, the contour regions for the mass of the heavier scalars $m_{H_i}$ ($i=1,2$)  in the plane $\tilde{m}_{u1}\hspace{0.2cm}vs\hspace{0.2cm}\tilde{m}_{d}$, assuming again  $m_{stop} = 1\hspace{0.1cm}TeV$. 
Our choice for the remaining parameters is the same as the one for $m_h$ (for fig 1a), which is also assumed 
 to evaluate  $m_{H_4}$, but in this case the mass formulae includes  one extra free parameter: $\tilde{m}_{u2}$, and we choose to plot $m_{H_4}$ precisely as a function of $\tilde{m}_{u2}$ (figure 1d).  These parameters define our benchmark point, which not only produces the value
$m_h \simeq 125$ GeV, but it also achieves masses  for the heavy Higgs bosons $H_2, H_3, H_4$) larger than about $O(0.5)$ TeV.  In fact, we notice from figures 1b,c,d, that the masses of the heavier Higgs bosons  can lay in the range:  (400 GeV-1TeV), which could be an interesting target for the HL-LHC.

\begin{figure}[H]
	\centering
	\subfigure[]{	\includegraphics[scale=0.22]{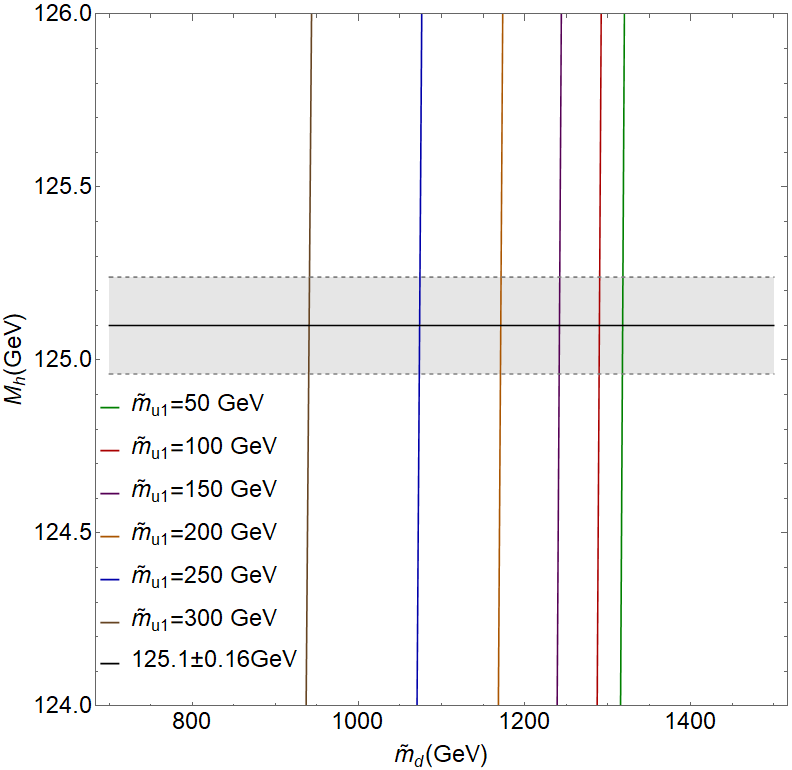}}
	\subfigure[]{     \includegraphics[scale=0.55]{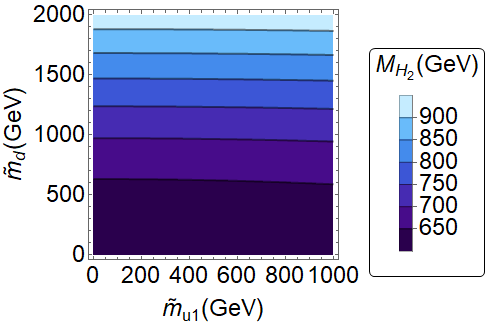}}
	\subfigure[]{      \includegraphics[scale=0.55]{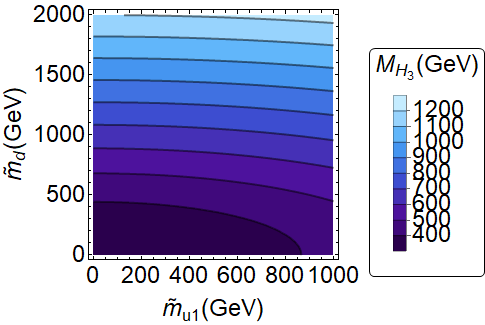}}
	\subfigure[]{     \includegraphics[scale=0.31]{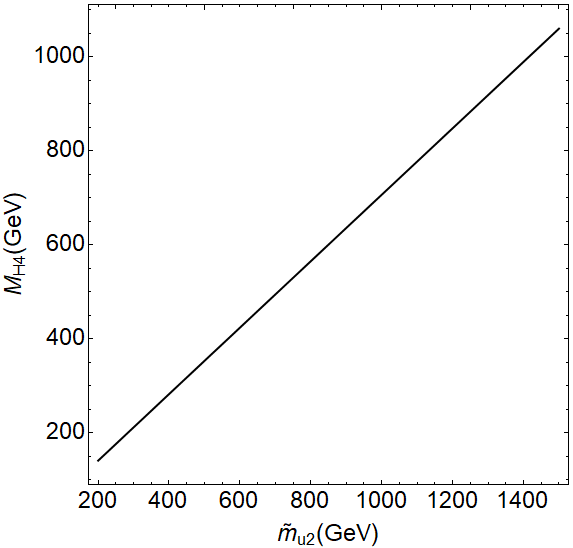}}

\caption{(a) The values of  $m_{h}$ as a function of  $\tilde{m}_{u1}$, (b) Contour regions of $m_{H_2}$ in the plane $\tilde{m}_{u1}- \tilde{m}_{d}$, (c) Contour regions of $m_{H_3}$ in the plane $\tilde{m}_{u1}- \tilde{m}_{d}$, (d) The values of $m_{H_4}$ as a function of $\tilde{m}_{u2}$. 
	}	
\end{figure}


\section{The Yukawa Lagrangian and FCNC in the Up quark sector}

As discussed in the motivations, in our model we assign one Higgs doublet to each fermion type, i.e each fermion type has its ``Private Higgs".
Thus,  the down quarks and the charged leptons only couple with a single Higgs, whereas the up quark sector could have couplings with two Higgs doublets simultaneously, $H_{u1}$ and $H_{u2}$. 

\subsection{Yukawa Lagrangian}

The Yukawa Lagrangian for this model can be derived from the corresponding superpotential, after eliminating the auxiliary fields (F-terms), 
in the end it can be written as follows
\begin{equation}
\mathcal{L}=\bar{u}_{iL}u_{jR}(Y^u_1)_{ij}H^0_{u1}+\bar{u}_{iL}u_{jR}(Y^u_4)_{ij}H^0_{u2}+\bar{d}_{iL}d_{jR}(Y^d)_{ij}H^0_d+\bar{l}_{iL}l_{jR}(Y^l)_{ij}H^0_l+h.c.
\end{equation}
Unlike the Up sector that interacts with two Higgs doublets at the same time, the down type quarks and the charged leptons only interact with a single one, implying that the corresponding mass matrices are given by
\begin{equation}
M_f=Y ^f \frac{v_f}{\sqrt{2}},
\end{equation}
where $f = d, l$, respectively.\\

After rotating to the mass eigenstate basis, both charged leptons and down type quarks Yukawa matrices become diagonal, i.e.
\begin{equation}
\bar{Y}^f=\frac{\sqrt{2}}{v_f}\bar{M}_f.
\end{equation}

On the other hand both Higgs doublets $H_{u1}$ and $H_{u2}$, couple with up-type fermions through the Yukawa matrices $Y^u_1$ and $Y^u_4$. After Spontaneous Symmetry Breaking (SSB), these matrices combine to produce a fermion mass matrix with some structure. The corresponding mass matrix receives contributions from both vevs $v_1$ and $v_4$, i.e.:
\begin{equation}
M_u=\frac{1}{\sqrt{2}}\left(v_1 Y^u_1+v_4 Y^u_4\right).
\end{equation}
To obtain physical fermion masses we need to diagonalize the mass matrix; this is achieved through a bi-unitary
transformation $\mathcal{V}_{L,R}$, i.e.
\begin{equation}
\bar{M}_u= \mathcal{V}_L M_u\mathcal{V}^\dagger_R=\mathcal{V}_L \frac{1}{\sqrt{2}}\left(v_1 Y^u_1+v_4 Y^u_4\right) \mathcal{V}^\dagger_R,
\end{equation}
the form of the matrix $\mathcal{V}_{L,R}$ depends on the texture type of $M_u$; closed expressions  have been obtained for the 4- and 6-texture hermitian and non-hermitian cases. Although $\mathcal{V}_{L,R}$ diagonalizes the matrix $M_u$, it does not necessarily diagonalize each of the Yukawa matrices that build up $M_u$, thus neutral flavor violating Higgs-fermion interactions will be induced in principle.
\\
\subsection{FCNC Higgs interactions and  Up-type quarks mass matrix  with 4-texture type}

The use of textures within the context of the 2HDM, considered first a specific form with six- zeroes \cite{tx1}, and it was found 
that the textures implied a pattern for the Higgs-fermion couplings  of the form $\frac{\sqrt{m_im_j}}{v}$, known as the Cheng-Sher ansatz. 
It turns out that such vertex satisfies limits on FCNC mediated by the Higgs bosons, with masses lighter than $\mathcal{O}$(TeV). 
The case with four-zero textures was presented in \cite{tx2,tx3}, and its implications were studied in \cite{tx4, tx5,tx6,DCT}. 
Other variations for the  Yukawa matrices were discussed in  \cite{tx7,tx8,Arroyo:2013tna}.\\

For completeness we include here the form of  the  four-zero texture matrix, namely:

\begin{equation}
M_u=
\begin{pmatrix}
0 & D & 0\\
D*& C & B\\
0 & B*& A\\
\end{pmatrix}.
\end{equation}

Here we shall assume that the Yukawa matrices in the Up-sector ($Y^u_{1}, Y^u_{4}$) have this 4-texture structure, but
rather than focusing on some specific model, we shall consider  the general features of this case. We invoke this mass texture,
in order to have under control the FCNC Higgs interactions, but the phenomenology will be analyzed in terms of the Yukawa matrix
elements, in the mass-basis.

\bigskip

 The Lagrangian for the up-quarks  sector is
\begin{equation}
\mathcal{L}_u=\bar{u}_{iL}u_{jR}(Y^u_1)_{ij}H^0_{u1}+\bar{u}_{iL}u_{jR}(Y^u_4)_{ij}H^0_{u2}+h.c.,
\end{equation}
After substituting $H^0_{u1}$ and $H^0_{u2}$, the Lagrangian becomes:
\begin{equation}
\mathcal{L}_u=\bar{u}_{iL}u_{jR}(Y^u_1)_{ij}\frac{1}{\sqrt{2}}(v_1+\eta_1+i\chi_1)+\bar{u}_{iL}u_{jR}(Y^u_4)_{ij}\frac{1}{\sqrt{2}}(v_4+\eta_4+i\chi_4)+h.c.,
\end{equation}
then keeping only the real part and factorizing, one obtains
\begin{equation}
\begin{split}
\mathcal{L}_u &=\bar{u}_L\left[\frac{1}{\sqrt{2}}(v_1Y^u_1+v_4Y^u_4)\right]u_R+\bar{u}_L\left[\frac{1}{\sqrt{2}}(\eta_1Y^u_1+\eta_4Y^u_4)\right]u_R+h.c.,\\
&=\bar{u}_LM_u u_R+\bar{u}_L\left[\frac{1}{\sqrt{2}}(\eta_1Y^u_1+\eta_4Y^u_4)\right]u_R+h.c.
\end{split}
\end{equation}

Then, one can express the rotated mass matrix as follows:  
\begin{equation}
\bar{M_u}=V_L M_u V_R^\dagger=\frac{v_1}{\sqrt{2}}V_LY^u_1V^\dagger_R+\frac{v_2}{\sqrt{2}}V_LY^u_4V^\dagger_R=\frac{v_1}{\sqrt{2}}\tilde{Y}^u_1+\frac{v_2}{\sqrt{2}}\tilde{Y}^u_4
\end{equation}
where $V_{L,R}$ denote the diagonalizing matrices in the up-quark sector, and $\tilde{Y}^u_{1,4}$ are the rotated Yukawa matrices 
in the mass-basis. Then, in the mass basis, the Yukawa lagrangian becoms:
\begin{equation}
\mathcal{L}_u=\bar{u}_L\bar{M}_u u_R+\bar{u}_L\left[\frac{1}{\sqrt{2}}(\eta_1\tilde{Y}^u_1+\eta_4\tilde{Y}^u_4)\right]u_R+h.c.,
\end{equation}


One can then express one of the Yukawa matrices in terms of the other one, and the diagonal mass matrix, namely:
\begin{equation}
\tilde{Y}^u_1=\frac{\sqrt{2}}{v_1}\bar{M}_u-\frac{v_4}{v_1}\tilde{Y}^u_4,
\end{equation}
notice that $\bar{M}_u$ is diagonal while $\tilde{Y}^u_i$ are not, with $i=1,4$.\\
Rewriting the Lagrangian in terms of $\tilde{Y}^u_{4}$, one  obtains:
\begin{equation}
\mathcal{L}_u=\bar{u}_L\bar{M}_u u_R+\frac{1}{\sqrt{2}}\bar{u}_L\left[\left(\frac{\sqrt{2}}{v_1}\bar{M}_u-\frac{v_4}{v_1}\tilde{Y}^u_4\right)\eta_1+\tilde{Y}^u_4\eta_4\right]u_R+h.c.
\end{equation}
Looking the last equation, we can see that now is only necessary to give information about the $\tilde{Y}^u_4$ matrix. Then we just need to express the neutral fields $\eta_{1,2}$ in terms of the Higgs mass eigenstates in order
to derive  the Higgs-fermion couplings in the up sector.

Thus, using Eq (\ref{eq:Diag})  we obtain the Higgs couplings in the up-type quark sector (with $c^{-1}_\omega=1/\cos \omega$),
\begin{align} \label{eq:hu}
g_{h_1u_iu_j}&=\frac{c_\omega c_{\delta _1}c_{\delta _2}}{vc_\beta}\left[\sqrt{2}(\bar{M}_u)_{ij}-vt_\omega s_\alpha c^{-1}_\omega (\tilde{Y}^u_4)_{ij}\right],\\
g_{h_2u_iu_j}&=-\frac{c_\omega(c_{\delta _3
	}s_{\delta _1}+c_{\delta _1}s_{\delta _2}s_{\delta _3} )}{vc_\beta}\left[\sqrt{2}(\bar{M}_u)_{ij}-vt_\omega s_\alpha c^{-1}_\omega (\tilde{Y}^u_4)_{ij}\right],\\
g_{h_3u_iu_j}&=\frac{c_\omega(s_{\delta _1}s_{\delta _3}-c_{\delta _1}c_{\delta _3} s_{\delta _2} )}{vc_\beta}\left[\sqrt{2}(\bar{M}_u)_{ij}-vt_\omega s_\alpha c^{-1}_\omega (\tilde{Y}^u_4)_{ij}\right],\\
g_{h_4u_iu_j}&=(\tilde{Y}^u_4)_{ij},
\end{align}

while for the down sector  one obtains:
\begin{align}\label{eq:hd}
g_{h_1dd}&=\frac{\bar{M}_d}{vs_\beta}c_\omega c_{\delta _2}s_{\delta _1},\\
g_{h_2dd}&=\frac{\bar{M}_d}{vs_\beta}c_\omega(c_{\delta _1}c_{\delta _3}-s_{\delta _1}s_{\delta _2}s_{\delta _3}),\\
g_{h_3dd}&=-\frac{\bar{M}_d}{vs_\beta}c_\omega(c_{\delta _3}s_{\delta _1}s_{\delta _2}+c_{\delta _1}s_{\delta _3}),
\end{align}

and finally for the lepton sector we have:
\begin{align}\label{eq:hl}
g_{h_1ll}&=\frac{\bar{M}_l}{vt_\omega c_\alpha}c_\omega s_{\delta _2},\\
g_{h_2ll}&=\frac{\bar{M}_l}{vt_\omega c_\alpha} c_\omega c_{\delta _2}s_{\delta _3},\\
g_{h_3ll}&=\frac{\bar{M}_l}{vt_\omega c_\alpha} c_\omega c_{\delta _2}c_{\delta _3}.
\end{align}

By expanding the covariant derivatives of  the Higgs doublets, we can also derive the $h_iVV$ couplings. 
The couplings  $h_iWW$ are expressed as follows
\begin{align}\label{eq:hW}
g_{h_1WW}&=2\frac{m^2_W}{v}c_\omega[c_\beta c_{\delta _1}c_{\delta _2}+ s_\beta c_{\delta _2}s_{\delta _1}+ t_\omega s_\alpha s_{\delta _2}]\\
g_{h_2WW}&=2\frac{m^2_W}{v}c_\omega[-c_\beta(c_{\delta _3}s_{\delta _1}+c_{\delta _1}s_{\delta _2}s_{\delta _3})+ s_\beta (c_{\delta _1}c_{\delta _3}-s_{\delta _1}s_{\delta _2}s_{\delta _3})+ t_\omega s_\alpha c_{\delta _2}s_{\delta _3}]\\
g_{h_3WW}&=2\frac{m^2_W}{v}c_\omega[c_\beta(s_{\delta _1}s_{\delta _3}-c_{\delta _1}c_{\delta _3}s_ {\delta _2})  -s_\beta(c_{\delta _3}
s_{\delta _1}s_{\delta _2}+c_{\delta _1}s_{\delta _3}) + t_\omega s_\alpha c_{\delta _2}c_{\delta _3}]\\
g_{h_4WW}&=2\frac{m^2_W}{v}c_\omega t_\omega c_\alpha
\end{align}
Similar results are obtained for the vertices $h_iZZ$.

Having all the relevant Higgs couplings, we are ready to work on the Higgs phenomenology. But before that,
we find interesting to comment about the MSSM limit of our model. In fact, this was discussed in Ref.
\cite{Aoki:2011yy}, which works in the so-called  Higgs-basis where only two doublets develop v.e.v.'s. 
The authors of this reference discusses in detail the issue of the decoupling limit, and although they confirm
that the model reduces to the SM, in the limit when all mass-parameters are very large, i.e. 
with no non-decoupling effects from the extra degrees of freedom, they also identify some
"quasi-decoupling" effects, which prevent the model to reduce to the MSSM in that limit. This is
 due to the mixing with the extra Higgs doublets, unless  the extra $b$-terms are set to zero, 
 such that no mixing is allowed.  On the other hand, using our parameterization for the v.e.v.'s, we can see that 
 in the limit $\alpha \to 0$ and  $\omega \to 0$, the leptons are massless, while the fourth Higgs doublet does not develop a v.e.v.. Furthermore, taking $\delta_2, \delta_3 \to 0$, we notice that the third and fourth neutral CP-even Higgs bosons ($H_3, H_4$) do not mix with the two lightest MSSM-like Higgs bosons $H_1 (=h)$ and  $H_2$. Thus this limit resembles the MSSM Higgs sector, but with massless leptons. 
 

\section{Model constraints from the LHC  Higgs searches}

Having analyzed the Yukawa and Higgs sector, we can focus now on the constrains of the 4HDM parameters involved in 
our study. Namely, we shall need to specify possible values for the following parameters:

\begin{itemize}
	\item Mixing angles $\delta_i$ ($i=1,\,2,\,3$) that appear in the rotation mass matrix; eq.\eqref{matrixO}.
	\item Angles that parameterize the VEV's: $\alpha,\,\beta,\,\omega$; eq. \eqref{eq:vev_2}.
	\item Heavy scalar masses: $m_{H_2}$, mainly.
	\item Yukawa matrix elements $\left(\tilde{Y}_4^u\right)_{tt}\equiv Y_{tt}$ and $\left(\tilde{Y}_4^u\right)_{tc}\equiv Y_{tc}$.
\end{itemize}  
One would also need to specify the soft SUSY-breaking terms, but they are essentially fixed by requiring the light Higgs boson mass
of 125 GeV, and the rest of the neutral CP-even Higgs masses to be larger than about 0.5 TeV.

In order to have a realistic scenario, we use the most up-to-date experimental measurements reported by ATLAS and CMS collaborations \cite{Aad:2019mbh, Sirunyan:2018koj}; namely, the signal strengths $\mathcal{R}_{X}$, which are define as follows:
\begin{equation}
\mathcal{R}_{X}=\frac{\sigma(pp\to h)\cdot BR(h\to X)}{\sigma(pp\to h^{\text{SM}})\cdot BR(h^{\text{SM}}\to X)},
\end{equation}
where $\sigma(pp\to H_i)$ is the production cross section of $H_i$, with $H_i=h,\,h^{\text{SM}}$; here $h$ is the SM-like Higgs boson coming from an extension of the SM and $h^{\text{SM}}$ is the SM Higgs boson; $BR(H_i\to X)$ is the branching ratio of $H_i$ decaying into a $X=b\bar{b},\;\tau^-\tau^+,\;\mu^-\mu^+,\;WW^*,\;ZZ^*,\;\gamma\gamma$. From the Higgs-fermion couplings (eq. \eqref{eq:hu}), we observe that the term outside the brackets must be close to the unity, while the second term inside the brackets must be close to zero, in order to have small derivations from the SM couplings. This is ensured by assuming that $c_{\omega}\sim c_{\beta}\sim c_{\delta_i}\sim 1$ and $s_{\alpha}\ll 1$.

\begin{figure}[h!]
\centering
 \includegraphics[scale=0.3]{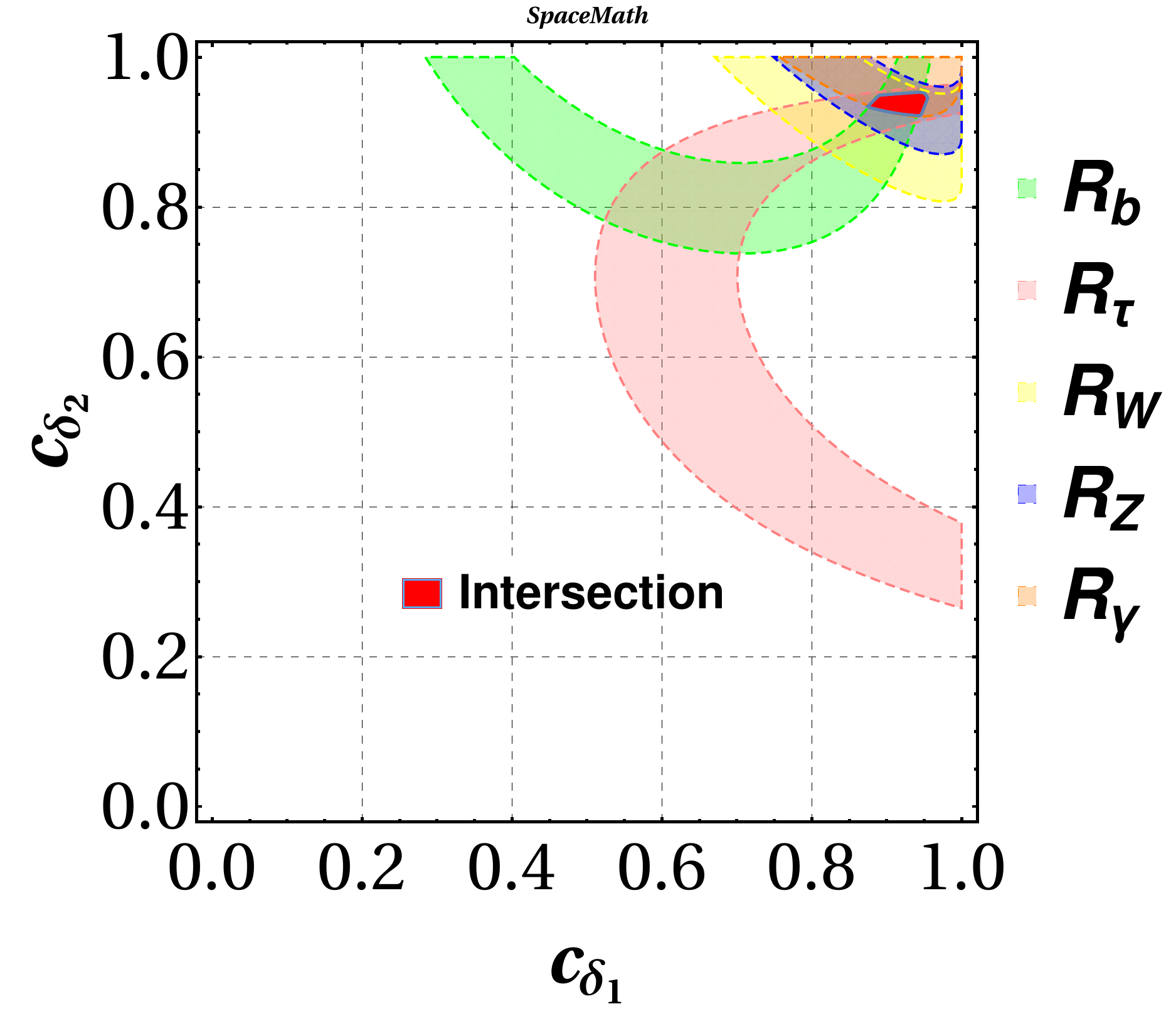}
\caption{Allowed region in the 
 plane $c_{\delta_1}-c_{\delta_2}$, from our analysis of the ratios  $\mathcal{R}_{b}$ (green), $\mathcal{R}_{\tau}$ (pink), $\mathcal{R}_{W}$ (yellow), $\mathcal{R}_{Z}$ (blue), $\mathcal{R}_{\gamma}$ (orange) in the 
 plane $c_{\delta_1}-c_{\delta_2}$. The red area represents the intersection of all individual allowed regions. }
\label{ParamSpace}
\end{figure}

Figure \ref{ParamSpace} shows the plane $c_{\delta_1}-c_{\delta_2}$, in which the filled areas represent the regios allowed  by $\mathcal{R}_{b}$ (green), $\mathcal{R}_{\tau}$ (pink), $\mathcal{R}_{W}$ (yellow), $\mathcal{R}_{Z}$ (blue), $\mathcal{R}_{\gamma}$ (orange). In turn, the intersection of all allowed regions is represented by the red area. The graph was generated by the $\texttt{SpaceMath}$ package \cite{Arroyo-Urena:2020qup}. We present in Table \ref{ParamValues} the values for the additional parameters used to find that plane.   
\begin{center}
\begin{table}

\caption{Values for additional parameters used to evaluate $\mathcal{R}_{x\bar{x}}$.\label{ParamValues}}

\begin{centering}
\begin{tabular}{cc}
\hline 
Parameter & Value\tabularnewline
\hline 
\hline 
$c_{\omega}$ & $0.93$\tabularnewline
\hline 
$s_{\alpha}$ & $0.01$\tabularnewline
\hline 
$c_{\beta}$ & $0.9$\tabularnewline
\hline 
$(\tilde{Y}^u_4)_{tt}$ & $0.1$\tabularnewline
\hline 
\end{tabular}
\par\end{centering}

\end{table}

\par\end{center}

We are also required to know values of matrix elements $Y_{tc}$ (in what follows we shall denote $Y_{tc}= (\tilde{Y}^u_4)_{tc}$), because it is a fundamental parameter in our analysis, since we are interested in a possible evidence of the flavor changing decays $H_2\to tc$ and $t\to ch$ whose couplings are proportional to $Y_{tc}$. In order to constraint it, we  shall consider first the high-energy constraints coming from LHC bounds
on rare top decay $t \to ch$, then we shall compare this with the low-energy constraints, in particular $D-\bar{D}$ mixing.

For the high-energy constraints, we use the direct upper limit on $BR(t\to ch)<1.1\times 10^{-3}$ \cite{PDG}; with this value we obtain a bound on $Y_{tc}$ of order 1, depending on the values of $c_{\beta}$. Nevertheless, the authors of the Ref. \cite{Papaefstathiou:2017xuv} have obtained an estimation by extrapoling the number of events for the signal and backgrounds from 36.1 fb$^{-1}$ to 3000 fb$^{-1}$, assuming that the experimental details and analysis remain unchanged. This resulting upper limit is given by $BR(t\to ch)<7.69\times 10^{-5}$. 

\begin{figure}[h!]
\centering
 \includegraphics[scale=0.2]{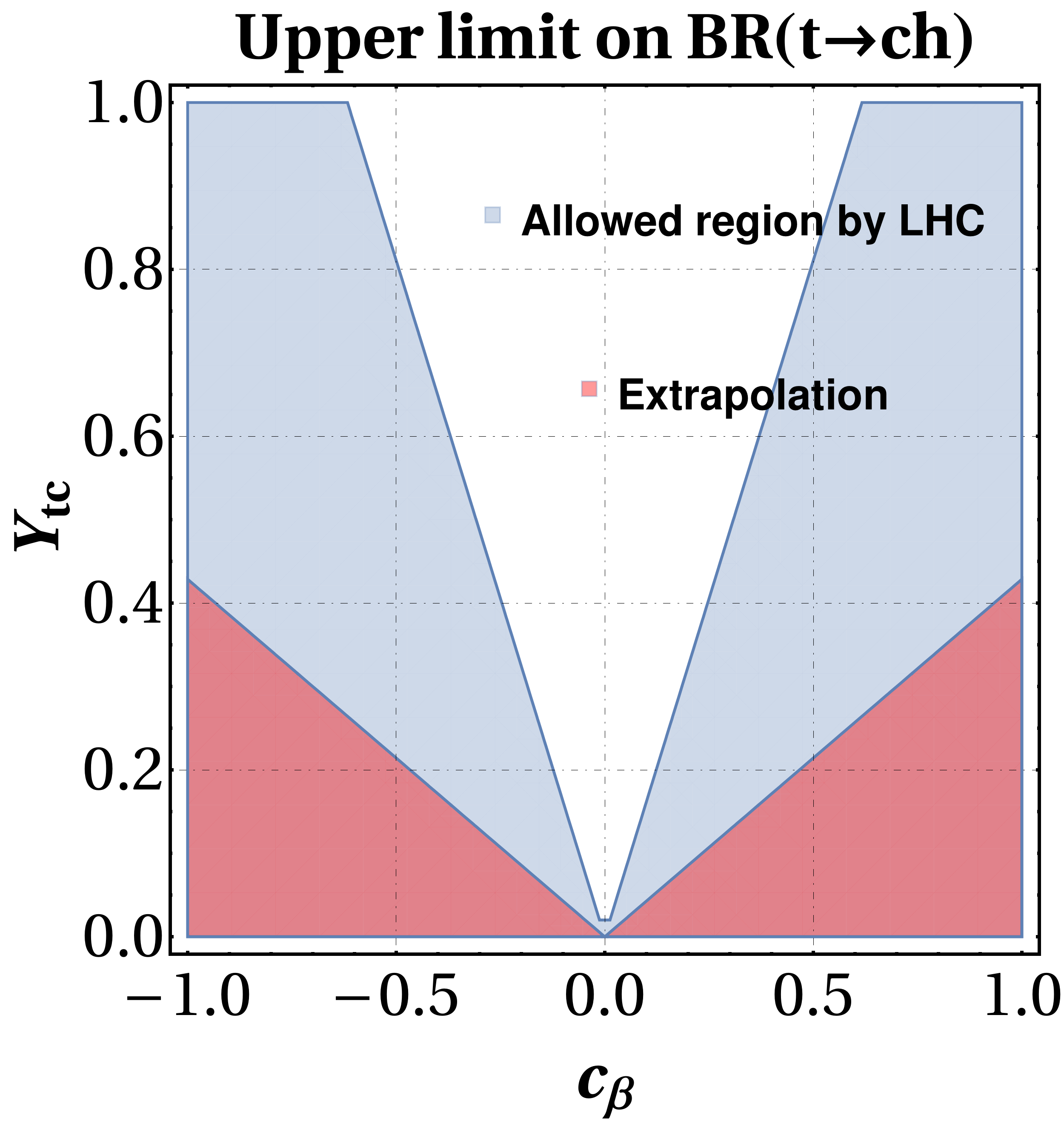}
\caption{Plane $c_{\beta}-Y_{tc}$. The filled areas represent the allowed regions from the direct upper limit reported by ATLAS collaboration on $BR(t\to ch)$ (blue region) and by extrapolation (red region).
 }
\label{BR_tch}
\end{figure}

Figure \ref{BR_tch} shows the allowed regions in the plane $c_{\beta}-Y_{tc}$, where the shaded areas represent the allowed regions obtained from the direct upper limit reported in Ref \cite{PDG}  on $BR(t\to ch)$ (blue area), and by the extrapolation analysis (red area). We note that, by considering $c_{\beta}=0.9$ (see Table \ref{ParamValues}), $Y_{tc}$ can reach a value of up to about 0.4, after taking into account  the extrapolation analysis.
 
Now, regarding the low-energy constraints, Ref. \cite{Babu:2018uik} has presented a detailed analysis of FCNC arising from
 the 2HDM incorporating  the Cheng-Sher anzast, where the couplings of the Higgs bosons ($H_a$) with fermions $f_i f_j H_a $ are of the form 
$ \chi^a_{ij}(m_i m_j)^{1/2}/v$, finding that the limits on the coefficients  $\chi^a_{ij}$ lay in the range $0.1-0.5$, 
which already requires some moderate fine-tuning.
 However, such analysis does not apply completely to our model, as in our case there are extra suppressions factors coming from the mixing angles. In particular, the low- energy-constraints on the parameter $Y_{tc}$, coming from the one-loop contribution  to the $D-\bar{D}$ mixing
 with neutral Higgs-exchange, were considered recently  in  Ref. \cite{Altunkaynak:2015twa,Crivellin:2013wna},  with further 
comments from  ref. \cite{Hou:2020ciy}, finding the bound on the product $ |\rho_{tc} \rho_{tu}| \leq 0.02$, with $\rho_{tc} \simeq g_{H_2tc}$, 
modulo some mixing-angle factor.  Within our model we  find that the coupling $H_2 tc$ is  about $g_{H_2tc} \simeq 10^{-4}$ 
for $Y_{tc} \simeq 0.1$, which satisfies this constraint. 

 Regarding the tree-level neutral Higgs contribution to $D-\bar{D}$ mixing, which is sensitive to $Y_{uc}$,
we have done an estimate of this effect for the  parameters (mixing angles) favored by the Higgs LHC data. We worked along the lines of 
the analysis presented in ref. \cite{Altmannshofer:2019ogm}, which was done for 2HDM-III and claims that no significant bound on 
$Y_{uc}$ is obtained for that model. We find that the contribution from the lightests neutral Higgs bosons  within our model
 ($h$ and $H_2$, with $m_{H_2}=400$ GeV)  are bellow the experimental uncertainties, even for $Y_{uc} = 1.5$.

\bigskip
Thus, our model parameters satisfy all low-energy constraints, and in fact the strongest constraints will come from 
the LHC studies.
Table \ref{ParamValues} summarizes the benchmark point for the parameters from our model, to be used in the subsequent calculations. Besides, we also set: $c_{\delta_1}=c_{\delta_2}=0.95$, unless stated otherwise.


\section{The LHC search for the FCNC decays $t\rightarrow ch$, $H_2 \rightarrow tc$}

Top quark rare decays have been studied for several years as a channel to search for new physics \cite{Eilam:1990zc,DiazCruz:1989ub,DiazCruz:2001gf}, 
including a variety of theoretical calculations for $BR(t\to ch)$. As discussed before, our model allows for FCNC couplings in the up sector, therefore we can obtain a prediction for both the FCNC top quark and $H_i$ decays. 
On the other side, Refs. \cite{AguilarSaavedra:2004wm, Arroyo-Urena:2019qhl, Bolanos:2019dso}, provides some estimates for the branching ratios for $t\rightarrow ch$ that could be proved at the different phases of LHC. For instance, it is claimed that the top decay processes provide the best channel to discover top FCNC interactions, while only in some cases it is surpassed by single top production, when up and charm quark interactions are involved. In some of the examples discussed in Ref. \cite{AguilarSaavedra:2004wm}, it gives the maximum rates predicted to be observable with a 3$\sigma$ statistical significance or more, for one LHC year with a luminosity of 6000 fb$^{-1}$.

In the coming subsection we shall perform a detailed study of the detection of the decay $t\to ch$ at the coming High-luminosity phase of the LHC (HL-LHC), and then we shall present an analysis to determine the viability of the LHC to detect the decay $H_2\to tc$ at HL-LHC.

\subsection{Search for the decay $t\to c h$ at the LHC}

The branching ratio for the decay $t\to c h$, at tree level, can be to computed through the following expression:

\begin{equation}
 BR(t\rightarrow c h)=\frac{\Gamma(t\rightarrow ch)}{\Gamma_{tot}},
\end{equation}

where the total top width is given by: $\Gamma_{tot}=\Gamma(t\to Wb)+\Gamma(t\to ch)$  GeV, and the width for the FCNC top decay is:

\begin{equation}
 \Gamma(t\rightarrow ch)=\frac{m_t}{16\pi}g_{htc}^2\left[(1+r_{tc})^2-r_{ht}^2\right]\times\sqrt{1-(r_{ht}+r_{tc})^2}\sqrt{1-(r_{ht}-r_{tc})^2},
\end{equation}

here, $g_{htc}$ is given by eq.\eqref{eq:hu}, $r_{ht}=m_{h}/m_{t}$, $r_{tc}=m_{c}/m_{t}$and $v=246$ GeV.

\begin{figure}[h!]
\centering
 \includegraphics[scale=0.2]{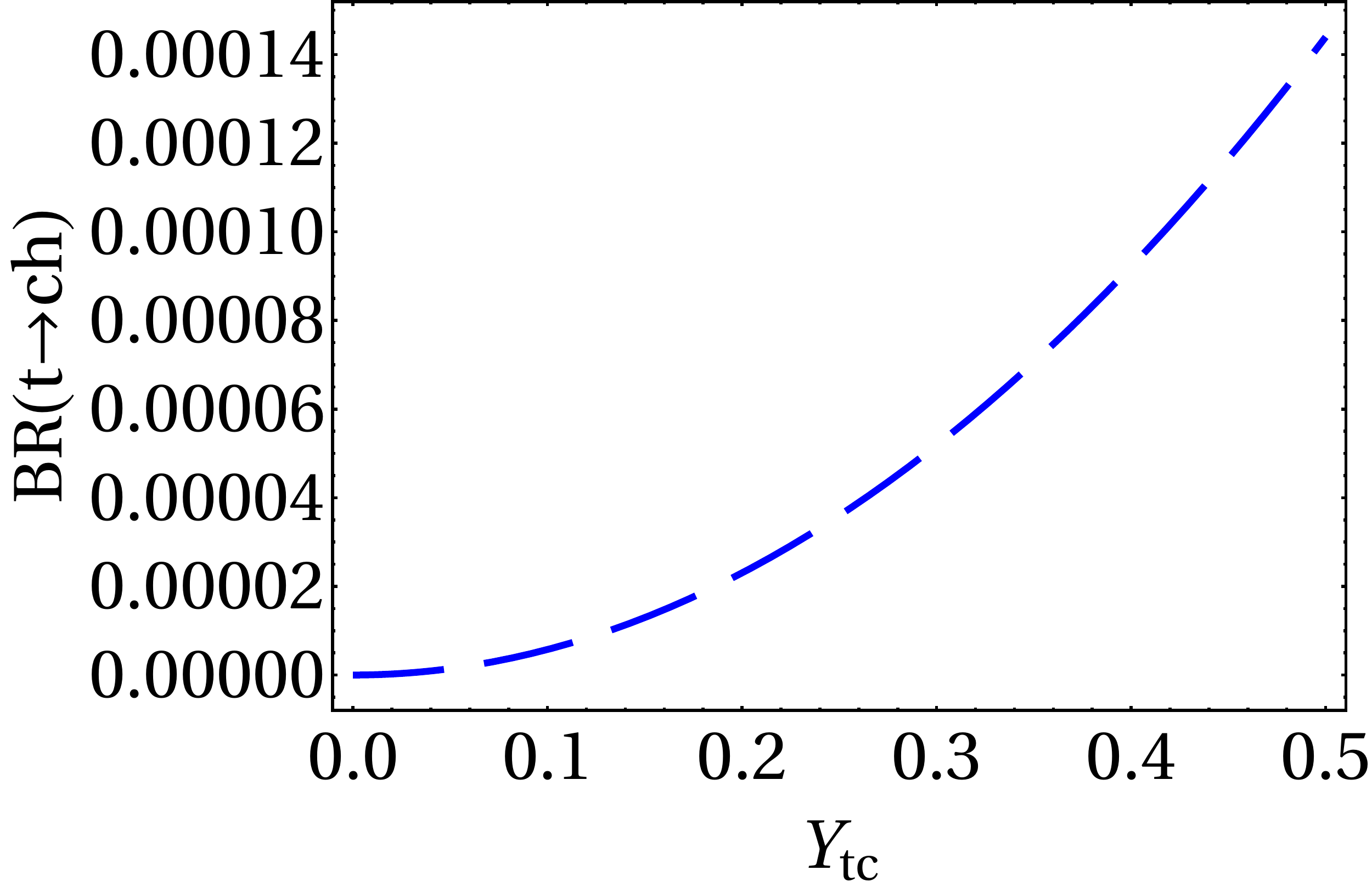}
\caption{Branching ratio of the decay $t\to ch$ as a function of $Y_{tc}$
\label{BR_tch1}
 }
\end{figure}

We display in Fig. \ref{BR_tch1} the $BR(t\to ch)$ as a function of $Y_{tc} $, 
here we use values for the additional parameters as shown in Table \ref{ParamSpace}. 
We observe that the top FCNC branching ratio can reach values of the order $10^{-4}$, which is very promising and motivates to undertake our study of detectability of the signal  at the future stages of the LHC. \\


The analysis is carried out for the LHC and its next stage, i.e, HL-LHC \cite{Apollinari:2017cqg}. Let us first discuss both the signal and main SM backgrounds, for the decay channels of the Higgs boson to be considered here. We adopt the strategy carried out by ATLAS and CMS collaborations \cite{Aaboud:2018pob, Sirunyan:2017uae}.
\begin{itemize}
	\item \textbf{Signal} 

		We consider top pair production, and then one top decays via the FCNC mode, while the other one decays through the SM mode. Thus, the signature is $pp\to t\bar{t}\to hc+Wb\to Xc+\ell\nu_{\ell}b$, and we shall consider the Higgs decays:  $X=\gamma\gamma$ or a $b\bar{b}$ pair. Thus, we identify the final state by the modes:  $\gamma\gamma bj\ell\nu_{\ell}$ or $b\bar{b}bj\ell\nu_{\ell}$.
	
	\item \textbf{Background}   
	\begin{enumerate}
		\item In the case of the \textit{diphoton-channel}, we study the backgrounds coming from: 
		\begin{itemize}
			\item $pp\to t\bar{t}h$,
            \item $pp\to hjjW^{\pm}$,
            \item $pp\to t\bar{t}\gamma\gamma$,
            \item $pp\to \gamma\gamma jj W^{\pm}$.
		\end{itemize}
	\item For the mode: $\textit{bb-channel}$, we shall include the background from:
	\begin{itemize}
		\item $pp\to t\bar{t}\to b\ell^+\nu\bar{b}\bar{c}s+X$ or $pp\to t\bar{t}\to bc\bar{s}\bar{b}\ell^-\bar{\nu}+X$ with a $c$-jet mis-identified as a $b$-jet,
		\item $pp\to t\bar{t}\to b\ell\nu\bar{b}u\bar{d}$,
		\item $pp\to b\bar{b}b\bar{b}\ell\nu$,
		\item $pp\to b\bar{b}c\bar{c}\ell\nu$.
	\end{itemize}
	\end{enumerate}
\end{itemize}  

On the other hand, as fas as our computation scheme is concerned, we firts implement the Feynman rules of the model via $\texttt{LanHEP}$ routines for $\texttt{MadGraph5}$ \cite{Alwall:2011uj} and $\texttt{CalcHEP}$ \cite{Belyaev:2012qa}. In this way, the signal and background events are generated by $\texttt{MAdGraph5}$ interfaced with $\texttt{Pythia6}$ \cite{Sjostrand:2006za} and $\texttt{Delphes3}$ \cite{deFavereau:2013fsa} for detector analysis. We generate $10^5$ events for both the signal and background using the CT10 parton distribution functions \cite{Gao:2013xoa}.

\begin{figure}[h!]
	\centering
		\subfigure[ ]{\includegraphics[scale=0.15]{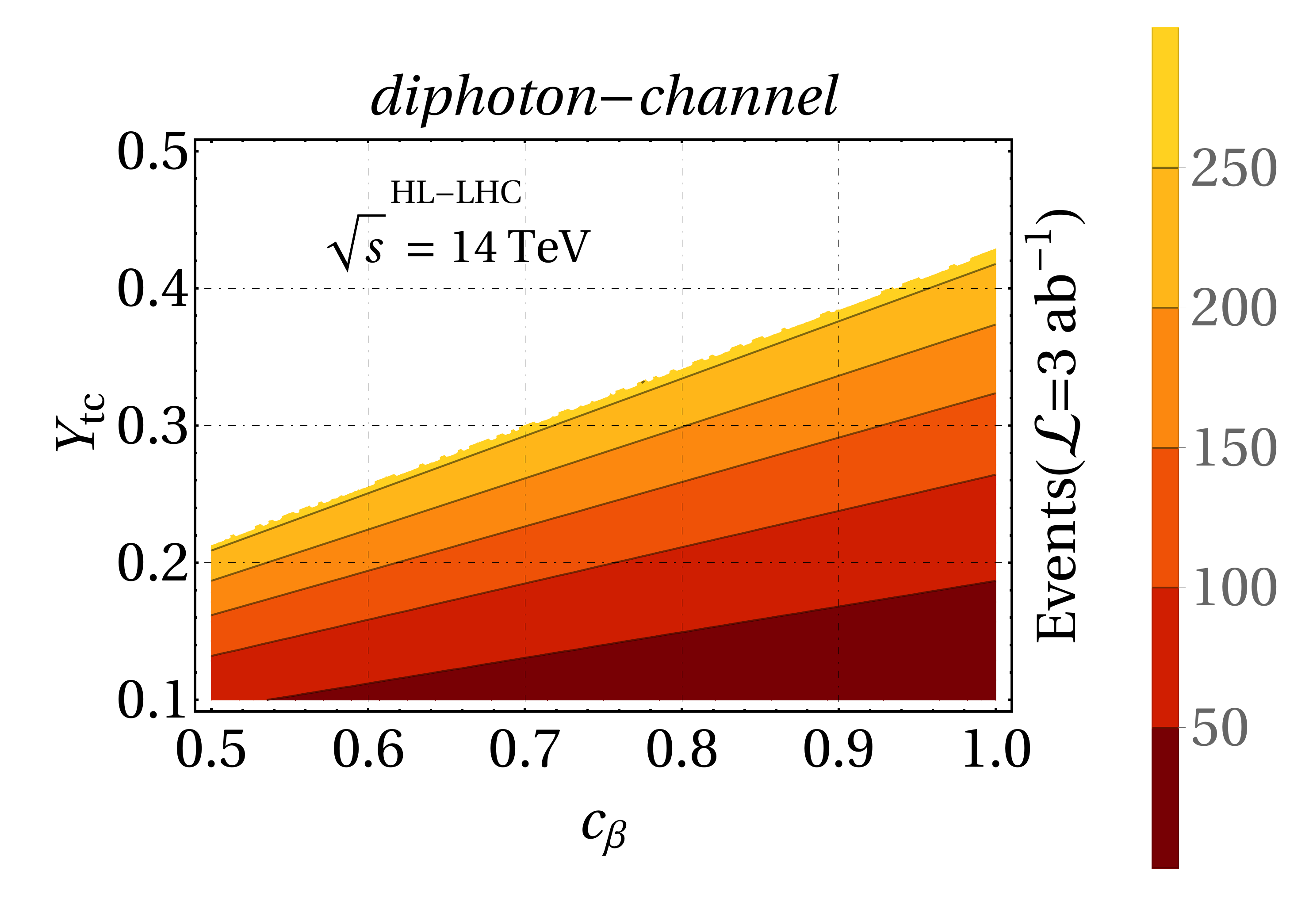}}
		\subfigure[ ]{\includegraphics[scale=0.15]{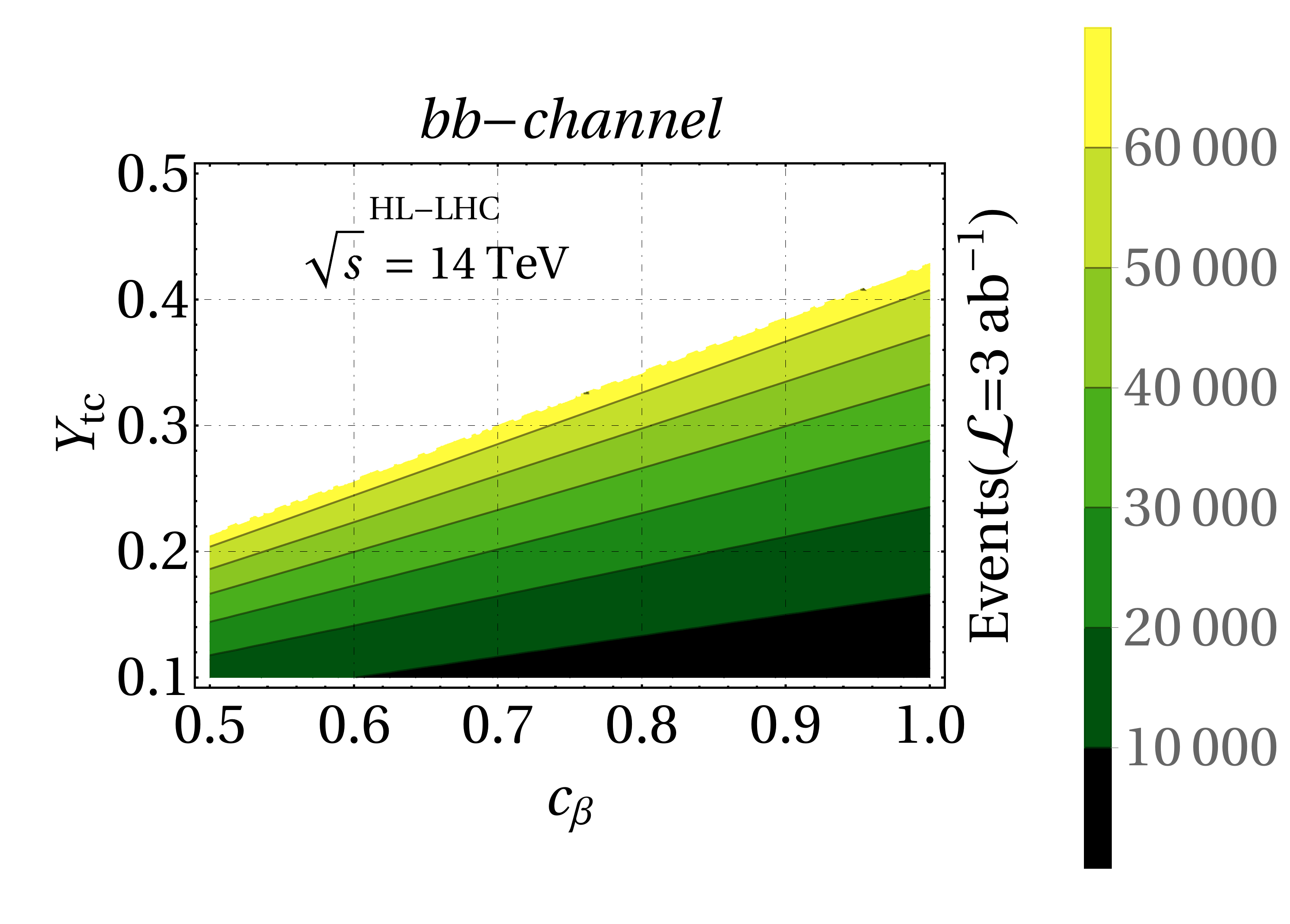}}
	\caption{Number of signal events as a function of $c_{\beta}$ and $Y_{tc}$ for (a) \textit{diphoton-channel} and (b) $\textit{bb-channel}$. In both cases we use the optimal integrated luminosity searched by the HL-LHC. We use the parameters shown in table 2.  \label{events_bb}}
\end{figure}

We now turn to evaluate the number of signal events produced as a function $c_{\beta}$ and $Y_{tc}$.
We present in Fig. \ref{events_bb} the signal events for the (a) \textit{diphoton-channel} and (b) $\textit{bb-channel}$  in the plane $c_{\beta}$-$Y_{tc}$ for HL-LHC, with an integrated luminosity of $3$ ab$^{-1}$. We observe that the number of signal events produced at the HL-LHC 
would be of order 250 (60000) for the \textit{diphoton-channel} ($\textit{bb-channel}$), assuming $Y_{tc}\sim 0.4$ and $c_{\beta}\sim 1$. For the LHC the number of signal events is about $0.1$ (one tenth) of the events expected HL-LHC.   

The ATLAS and CMS collaborations \cite{Aaboud:2018pob, Sirunyan:2017uae}, searched for the decay $t\to ch$ in both the \textit{diphoton-channel} and \textit{bb-channel}; nevertheless no significant deviation from the SM prediction was observed. We follow their strategies taking the same kinematic cuts for our analysis. These cuts are:

\begin{itemize}
\item  \textit{diphoton-channel}
\begin{enumerate}
\item We require exactly one $b-$jet and two photons.
\item We identify charged leptons and photons emerging from the signal imposing $p_T^{\gamma,\,\ell}>25$
GeV.
\item The main variable for search the Higgs boson decay into \textit{diphoton} system is the invariant mass, $M_{\gamma\gamma}$, which is chosen to lay between \textcolor{blue}{$120\leq M_{\gamma\gamma}\leq 130$ GeV.} 
\item Due to the Higgs boson decays into \textit{diphoton} system, it is required that the invariant mass associated to the top quark to fall between $160\leq M_{\gamma\gamma j}\leq 190$ GeV. 
\item The separations between the photons coming from Higgs boson decay must be $1.8<\Delta R_{\gamma,\,\gamma}<5.0$.
\item The separation between \textit{diphoton} system and the jet: $\Delta R_{\gamma\gamma,\,j}<1.8$.
\item \textcolor{blue}{Due to the non detected neutrino in the final state we demand a missing transverse energy $\slashed E_T>30$ GeV.}
\item The tagging and mistagging efficiencies chosen are as follows:
\begin{itemize}
\item $\epsilon_b=70\%$,
\item $\epsilon_c=14\%$,
\item $\epsilon_j=1\%$.
\end{itemize}
\end{enumerate}
\end{itemize}

\begin{itemize}
	\item  \textit{bb-channel}
	\begin{enumerate}
		\item We require exactly four jets: three of them are tagged as $b-$jets with $p_T^{j,\,b}>30$ GeV and $|\eta^j<2.5|$.
		\item Exactly one isolated lepton with: $p_T^{\ell}>20$ GeV and $|\eta^{\ell}| < 2.5$
		\item For having a neutrino emerging in the final state, the missing transverse energy $\slashed E_T>30$ GeV is required.
		\item To reconstruct the top quarks mass associated to flavor-changing neutral current, it is required that $|M_{b_1b_2j}-m_t|\leq 26$ GeV.
		\item As far as the reconstruction of the Higgs boson mass is concerned, it is imposed that $|M_{b_1b_2}-m_h|\leq 0.15m_h$.
		\item It is required $\Delta R$ between each jet and charged lepton pairs to be $\sqrt{\Delta\phi^2+\Delta\eta^2}>0.4$. 
		\item The tagging and mistagging efficiencies chosen are as follows:
		\begin{itemize}
			\item $\epsilon_b=70\%$,
			\item $\epsilon_c=14\%$,
			\item $\epsilon_j=1\%$.
		\end{itemize}
	\end{enumerate}
\end{itemize}

\begin{figure}[!htb]
	\centering
	\subfigure[ ]{\includegraphics[scale=0.4,angle=270]{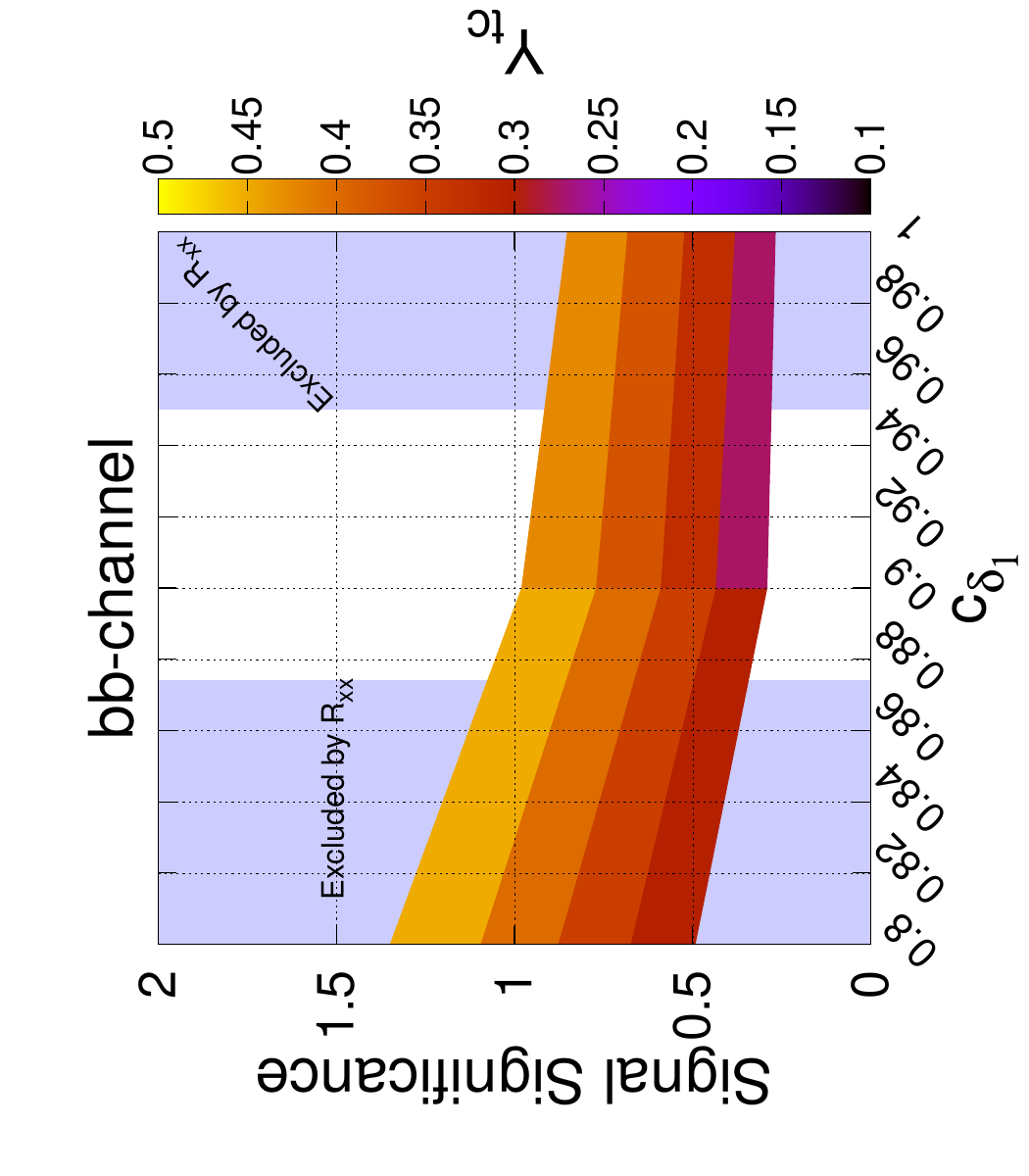}}
	\subfigure[ ]{\includegraphics[scale=0.4,angle=270]{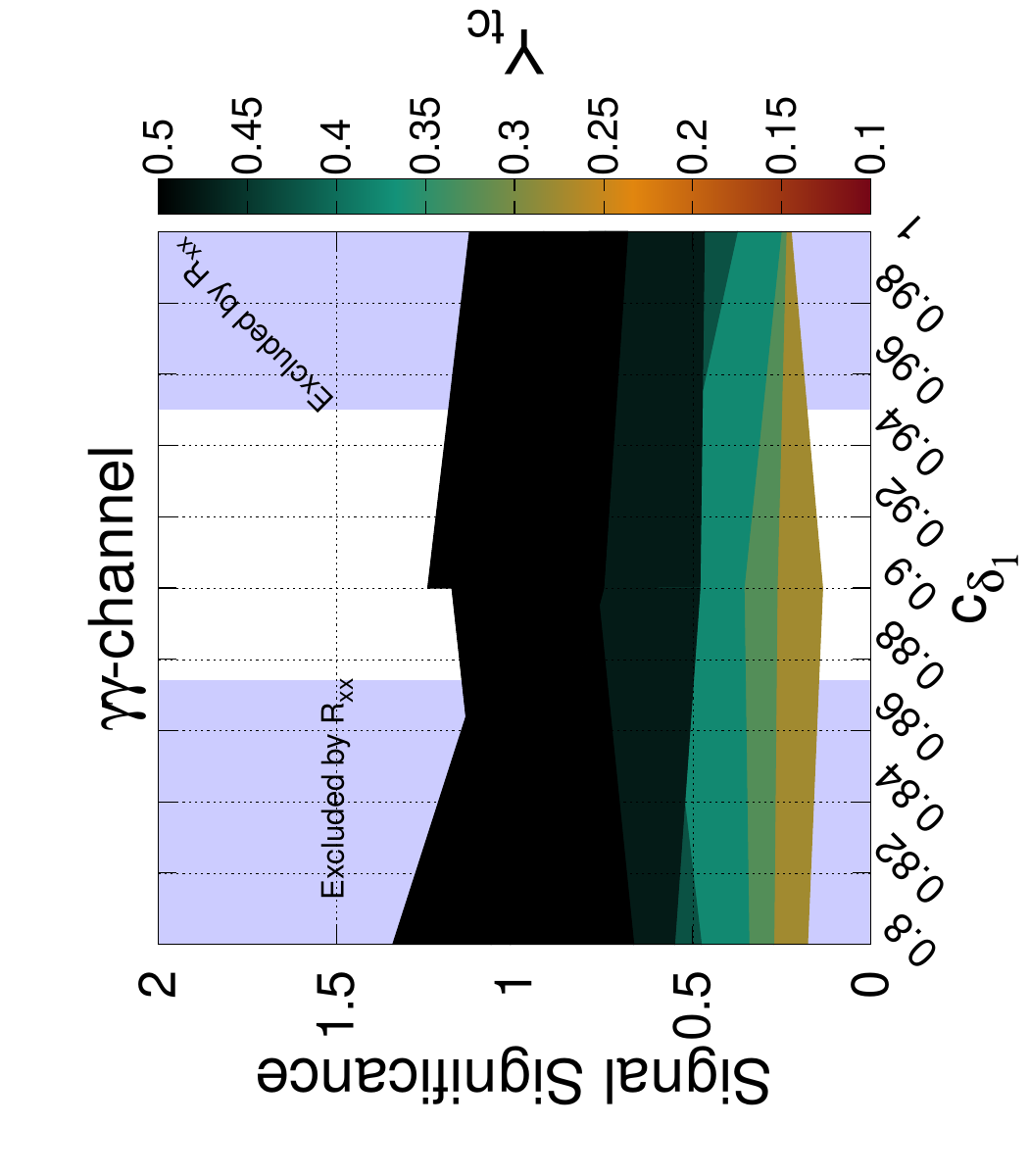}}
	\caption{Signal significance as a function of $c_{\delta_1}$ and $Y_{tc}$: (a)  \textit{bb-channel} and (b) \textit{diphoton-channel}. We set $c_{\delta_2}=0.95$, and the parameters shown in table 2.  Also shown the regions constrained by LHC Higgs data, as derived in the previous section. \label{SignalSignificancetch1}}
\end{figure}

We now turn to evaluate the signal significance  $\mathcal{S}=N_S/\sqrt{N_S+N_B}$, where $N_S$ is the number of signal events 
and $N_B$ is the number of background events, once the kinematic cuts were applied. 
Figure \ref{SignalSignificancetch1} shows the corresponding signal significance that can be achieved at the HL-LHC,
as a function of $c_{\delta_1}$ and $Y_{tc}$. 

We can appreciate that for both channels, values of the significance are of order 1 for $c_{\delta_1}$ within the range allowed by $\mathcal{R}_X$ and $Y_{tc}\sim \mathcal{O}(0.4)$. 

\subsection{Search for the decay $H_2\to tc$ at the LHC}

\begin{figure}[!htb]
 	\centering
 	\subfigure[ ]{\includegraphics[scale=0.22]{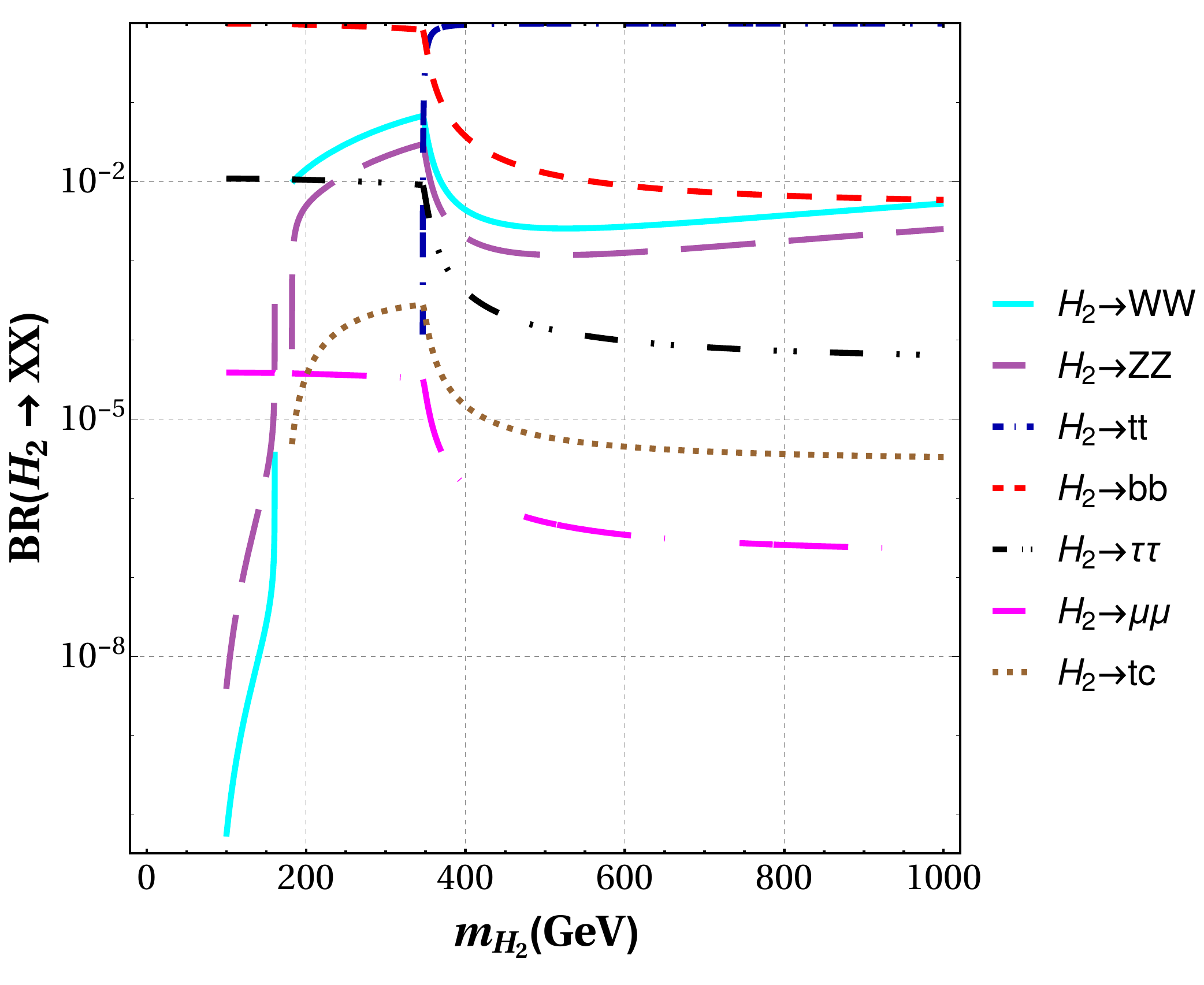}}
 	\subfigure[ ]{\includegraphics[scale=0.22]{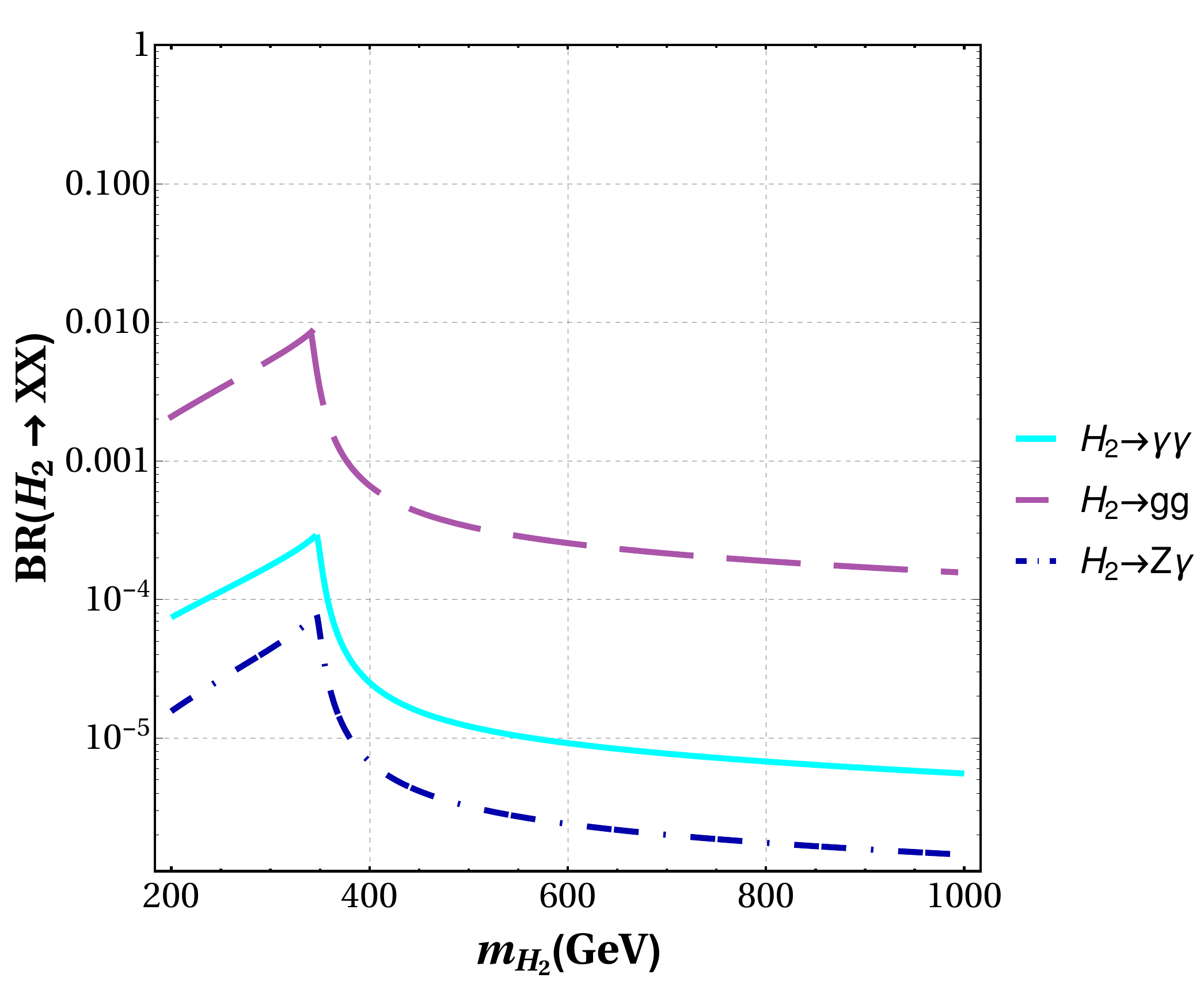}}
 	\caption{Branching ratio of the decays $H_2\to XX$: (a) tree level modes (a) one-loop level modes. We used the values of the parameters shown in Table \ref{ParamValues}. \label{BRs_h2}}
 \end{figure}

To begin with, we evaluate the relevant decay modes of $H_2$ into final states with two particles; the corresponding branching ratios are shown in Fig. \ref{BRs_h2}, with (a) modes at tree level and (b) modes at one-loop level.   

 We observe that the FCNC mode $H_2 \to tc$, reaches values of order $10^{-4}$, for a mass $m_{H_2} \sim 2 m_t$. Above this mass, this branching ratio reaches values of order $10^{-6}-10^{-5}$. The dominant decays modes for $m_{H_2}\leq 2m_t$, are into $bb$ and $VV$ ($V=\,W,\,Z$) pairs, whose branching ratios are of order $0.9$ and $10^{-2}$, respectively. Once the $tt$ channel is open, it becomes the dominant decay mode, with a branching ratio of order $1$. As far to the one-loop level decays, the mode $H_2\to gg$ has the largest B.R.
reaching values of order $10^{-2}$ ($10^{-3}$) for $m_{H_2}\leq2m_t$ ($m_{H_2}>2m_t$). The decay $H_2\to\gamma\gamma$ 
has B.R. of order $ 3 \times 10^{-4}-5\times 10^{-6}$, in the interval $2m_t-1000$ GeV. Finally,  we find
 $BR(H_2\to Z\gamma)$ can reach values as large as: $7\times 10^{-5}$. 

\begin{figure}[h!]
	\centering
	\subfigure[ ]{\includegraphics[scale=0.255]{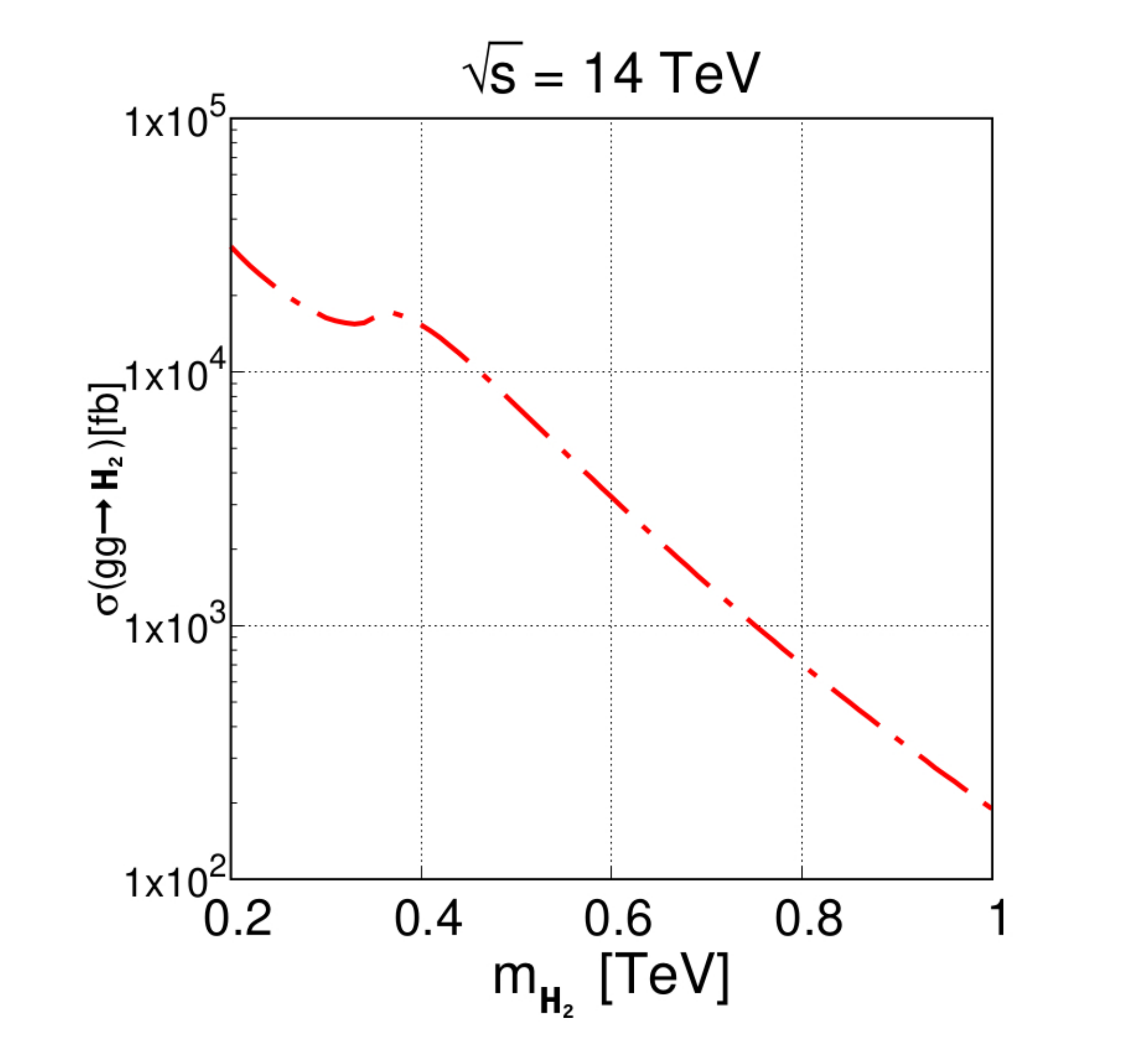}}
	\subfigure[ ]{\includegraphics[scale=0.28]{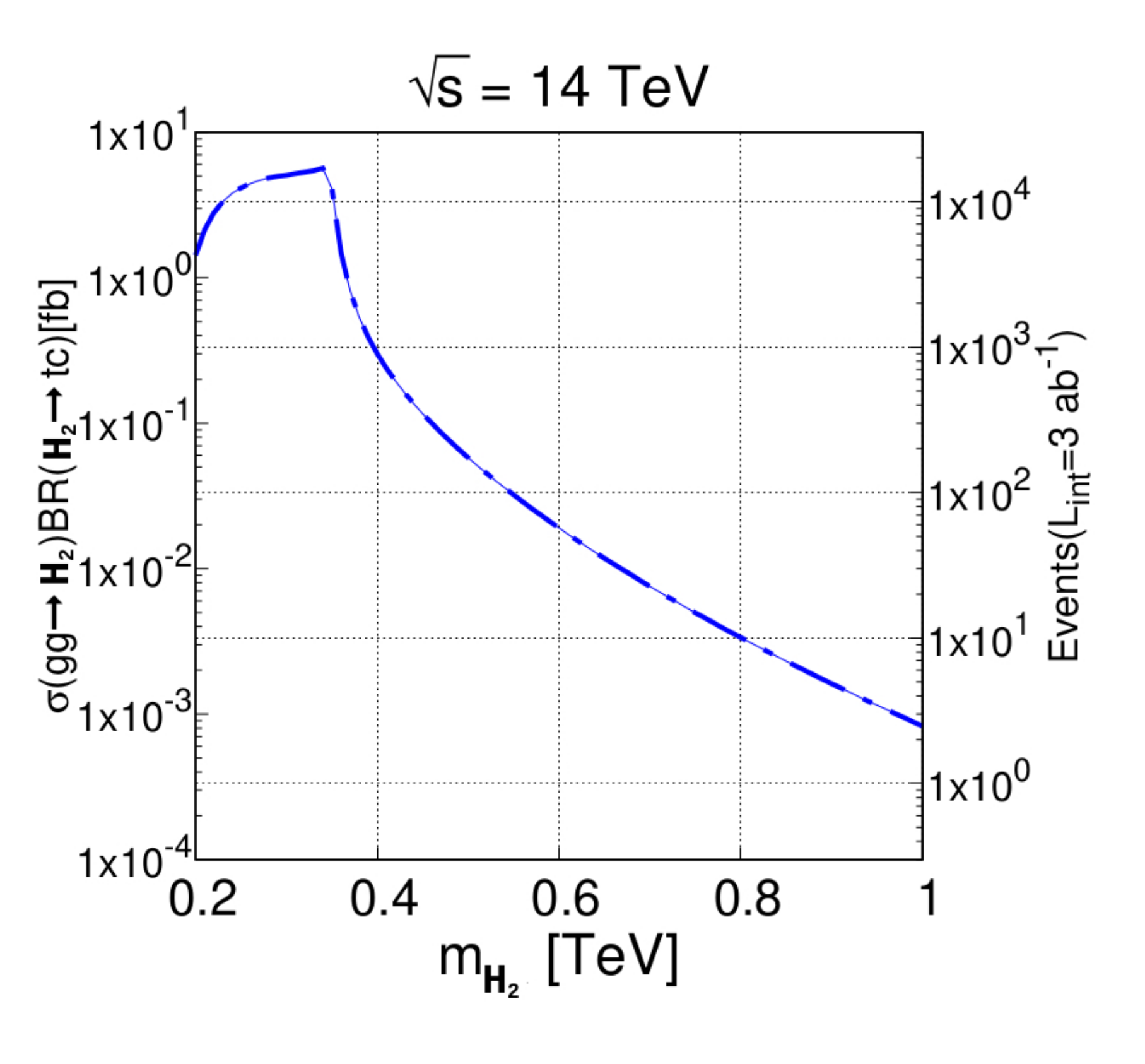}}
		\caption{(a) $H_2$ cross section as a function of $m_{H_2}$, through the gluon fusion mechanism and (b) Number of signal events at a center-of-mass energy equal to 14 TeV and integrated luminosity of 3 ab$^{-1}$. We used the values of the parameters shown in Table \ref{ParamValues}.  \label{XSh2} }
\end{figure}

For the production of the heavy Higgs boson, we focus on the gluon fusion mechanism, 
which  is the dominant production mechanism for the SM case; the corresponding cross section is displayed in Fig. \ref {XSh2}(a), while the number of signal events, considering the optimal integrated luminosity ($\mathcal{L}=3$ ab$^{-1}$) to achieve at the HL-LHC, is shown in Fig. \ref {XSh2}(b).

On the other hand, as in the previous section, we first define both the signal and the main SM background processes as follows:
\begin{itemize}
\item \textbf{Signal:} 

The signature searched is $gg\to h_2\to tc\to b\ell\nu_{\ell}c$, where $\ell=e,\,\mu$.
\item \textbf{Background:} 

The dominant SM background processes to the final state $bj\ell\nu_{\ell}$ comes from:
\begin{enumerate}
\item $Wjj+Wb\bar{b}$
\item $s$ and $t$ channel single top $tb+tj$
\item Another important background is the $t\bar{t}$ production where one of the two leptons is missed for both top quarks decaying semileptonically, or two of the four jets era missed when only one of the top quarks decays decays semileptonically.
\end{enumerate}
\end{itemize}

\begin{figure}[h!]
	\centering
		\includegraphics[scale=0.4]{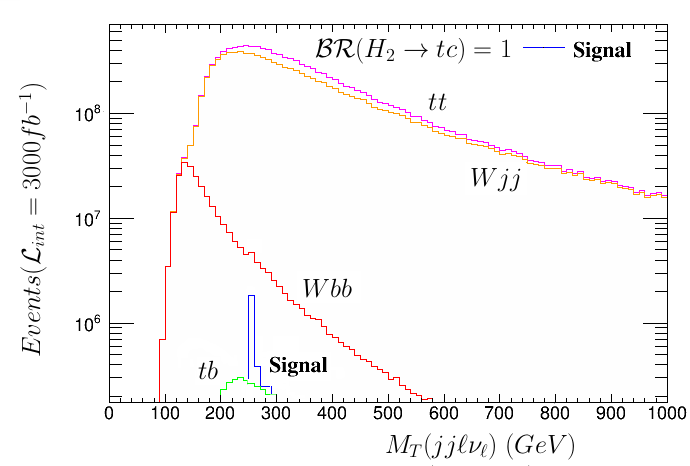}
		\caption{Transverse mass distribution without cuts. \label{MT}}
\end{figure}

The  kinematic cuts imposed are:
\begin{itemize}
\item The main kinematic cut to isolate the signal is the transverse mass $M_T$ which is defined as: 
\begin{equation}
M_T^{\ell}=\sqrt{2|\vec{P}_T^{\ell}||\vec{E}_T^{\text{miss}}|(1-\cos\Delta\phi_{\vec{P}_T^{\ell}-\vec{E}_T^{\text{miss}}})},
\end{equation}
Figure \ref{MT} shows the transverse mass distribution of the signal and backgrounds, without cuts, where we assume 
 $BR(H_2\to tc)=1$ in order to to highlight the signal. 
Then, we shall impose the cut:  $m_{H_2}-15<M_T^{\ell}<m_{H_2}+15$ (GeV).

\item  We require two jets with $|\eta^j|<2.5$ and $p_T^j>30$ GeV, one them is tagged as a $b$-jet.

\item  We require one isolated lepton ($e\,\text{or}\,\mu$) with $|\eta^{\ell}|<2.5$ and $p_T^{\ell}>20$ GeV.

\item We also consider a cut for the missing transverse energy $\slashed E_T>40$ GeV due to un-detected neutrinos.
\end{itemize}

 After applying the  above kinematic cuts  (and  also consider tagging and misstagging efficiencies as in previous section) 
 to the signal and the main background processes, we are able to compute the signal significance, defined  as $N_S/\sqrt{N_S+N_B}$,  where $N_S$ ($N_B$) are the number of signal (background) events, once the kinematic cuts were applied. 
 
\begin{figure}[h!]
	\centering
		\includegraphics[scale=0.4]{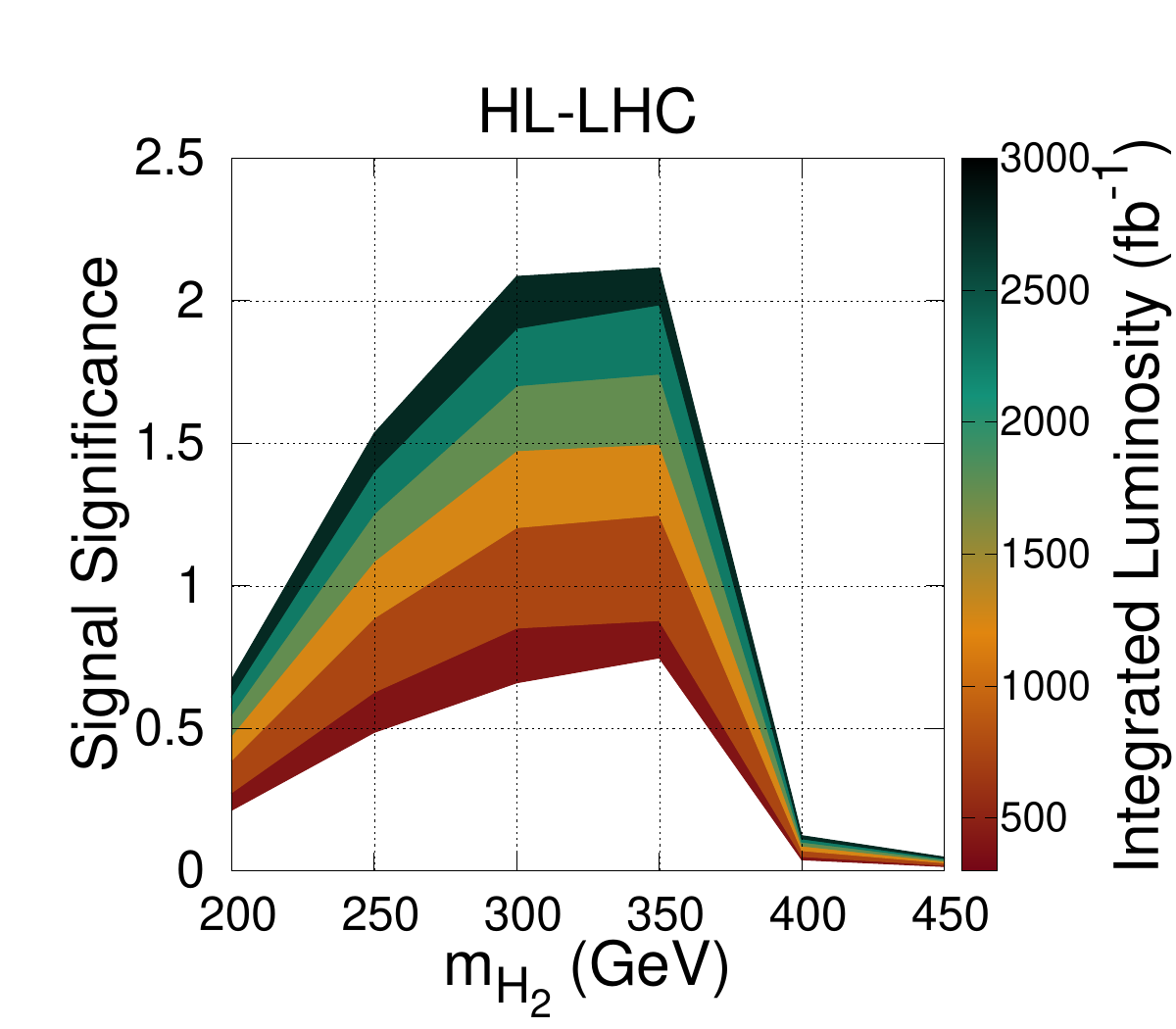}
		\caption{Signal significance as a function of the integrated luminosity and the $m_{H_2}$, using the parameters shown in table 2. \label{SignalSignificanceh2tc}}
\end{figure}

Meanwhile, we present in Fig. \ref{SignalSignificanceh2tc} the contours of the signal significance as a function of Higgs boson  mass 
$M_{H_2}$,  and displaying (in colors)  the integrated luminosity.

From this figure \ref{SignalSignificanceh2tc}, we can see that  the largest significance can be achieved with the largest
luminosity, for a certain  range of Higgs masses. For instance, we can  see that it will be possible to achieve a significance of  2.1$\sigma$ for a luminosity of order 3000 fb$^{-1}$ for $m_{H_2}\sim 2m_t$, while in the final stage of the LHC (with  an integrated luminosity of 300 fb$^{-1}$) 
the signal significance will  only be about  0.75$\sigma$ .

%
\section{Conclusions}

We presented a ``private" SUSY Higgs model that includes four Higgs doublets, where each fermion
type (up, down, and charged leptons) obtains its masses from a different Higgs doublet
$H_f$, with $ f = u1, d, e$, involving three out of the four Higgs doblets of the model. From the anomaly cancellation constraint, 
it turns out that the remaining doublet $H_{2u} $ could couple at most with up-type quarks, 
and thus in in principle this model could allow for the presence of  FCNCs in the up-quark sector sector. 
We studied the Yukawa Lagrangian and the Higgs potential of the model, in order to identify the Higgs mass eigenstates and
 their interactions. Then we identified the Yukawa couplings for the lighter scalars $H_1 (=h)$ and $H_2$. 
 Using the LHC results on the signal strengths,  we have derived constraints on the 
parameter space of our ``Private Higgs" model, including  those constrains obtained by LHC on the FCNC 
decay $t \rightarrow ch$. In general, the constraints on the FCNC parameters  ($\chi_{ij}$) of the 
 two-Higgs models of type-III, which are expected to be of $O(1)$, are now getting close to $O(0.1)$,
 which starts to signal the need of some fine-tunning for the consistency  of this type of models.

For the allowed region of parameter space, we have calculated the branching ratio for the decay $t \rightarrow ch$ ,
which reaches values of order $BR (t \rightarrow ch) \approx \mathcal{O}(10^{-4} - 10^{-5} )$. We then studied the prospects to 
improve on the detection of this mode at the future HL-LHC phase, focusing on the channels $t \to c h(\to \gamma\gamma)$ and
$t \to c h(\to b\bar{b})$, with a set of kinematical cuts inspired in the current searches of CMS and ATLAS. As a result
we find possible to cover such level of branching ratios, with a significances of slightly bellow  $1\sigma$  
for the channel $t \to c h(\to b\bar{b})$ with $Y_{tc} \simeq 0.3$, while the significance for the mode $t \to c h(\to \gamma\gamma)$ is 
slightly above $1\sigma$  for $Y_{tc} \simeq 0.3$. Clearly, more work is needed in order to find better ways to improve
the sensitivity of HL-LHC to the top quark decay $t \to ch$ through these Higgs decay modes.
 
 We also studied the FCNC decay of the next-to-lightest neutral CP-even
Higgs boson,  $H_2 \rightarrow tc$, which can also reach $BR(H_2 \rightarrow tc) \approx \mathcal{O}(10^{-4} - 10^{-5} )$. 
The detectability of the signal at  HL-LHC was also studied, imposing a set of standard cuts for both the signal and backgrounds, 
 finding that it could be possible to achieve a significance of   2.1$\sigma$ in the mass range $\simeq 300-350$ GeV, 
 with an integrated luminosity of  3000 fb$^{-1}$. 
 Our results are relevant because they can be seen as an extra motivation for the LHC experiments to improve  the search for the 
 FCNC top decays, and to explore the heavy Higgs spectrum through the FCNC decay modes.



\section*{Acknowledgments}
We would like to acknowledge the support of CONACYT and SNI. M. A. P\'erez de Le\'on is supported by a CONACYT graduate student fellowship. CIFFU is supported by VIEP-BUAP as a special project. B.L. thanks the support of UNAH.


\bigskip

\begin{thebibliography}{99}

\bibitem{lhc1} 

{Chatrchyan, S., Khachatryan, V., Sirunyan, A. M., Tumasyan, A., Adam, W., Aguilo, E., ... Friedl, M. (2012). Observation of a new boson at a mass of 125 GeV with the CMS experiment at the LHC. Physics Letters B, 716(1), 30-61.}\url{http://hdl.handle.net/1721.1/91933}

\bibitem{lhc2} 
{G. Aad et al. [ATLAS Collaboration], Observation of a new particle in the search for the Standard Model Higgs boson with the ATLAS detector at the LHC, Phys. Lett. B 716 (2012) 1 [arXiv:1207.7214 [hep-ex]]}\url{https://arxiv.org/abs/1207.7214}

\bibitem{lhc3} 
{G. Aad et al. [ATLAS and CMS Collaborations], Measurements of the Higgs boson production and decay rates and constraints on its couplings from a combined ATLAS and
 CMS analysis of the LHC pp collision data at $\sqrt{s} = 7$ and 8 TeV, JHEP 1608 (2016) 045
 [arXiv:1606.02266 [hep-ex]].}\url{https://arxiv.org/abs/1606.02266}

\bibitem{Delaunay:2013pja} 
C.~Delaunay, T.~Golling, G.~Perez and Y.~Soreq,
Phys.\ Rev.\ D {\bf 89}, no. 3, 033014 (2014)
doi:10.1103/PhysRevD.89.033014
[arXiv:1310.7029 [hep-ph]].

\bibitem{Arroyo-Urena:2020fkt} 
  M.~A.~Arroyo-Urena and J.~L.~Diaz-Cruz,
  arXiv:2005.01153 [hep-ph].


\bibitem{theprivate}
Porto, Rafael A.; Zee, A. The private higgs. Physics Letters B, 2008, vol. 666, no 5, p. 491-495.\url{http://arxiv.org/abs/arXiv:0712.0448}

\bibitem{privatelhc}
Bentov, Yoni; Zee, A. Private Higgs at the LHC. International Journal of Modern Physics A, 2013, vol. 28, no 28, p. 1350149. \url{https://arxiv.org/abs/1207.0467}

\bibitem{DCUS} 
Diaz-Cruz, J. L.,Salda\~na-Salazar, U. J. (2016). 
Higgs couplings and new signals from Flavon–Higgs mixing effects within multi-scalar models. Nuclear Physics B, 913, 942-963.\url{https://doi.org/10.1016/j.nuclphysb.2016.10.018}

\bibitem{Branco:2011iw} 
G.~C.~Branco, P.~M.~Ferreira, L.~Lavoura, M.~N.~Rebelo, M.~Sher and J.~P.~Silva,
Phys.\ Rept.\  {\bf 516}, 1 (2012)
doi:10.1016/j.physrep.2012.02.002
[arXiv:1106.0034 [hep-ph]].


\bibitem{Tsumura:2009yf}
K.~Tsumura and L.~Velasco-Sevilla,
Phys.\ Rev.\ D {\bf 81}, 036012 (2010)
doi:10.1103/PhysRevD.81.036012
[arXiv:0911.2149 [hep-ph]].


\bibitem{DCBG} 
Barradas-Guevara, E., Diaz-Cruz, J. L., Felix-Beltran, O., Saldana-Salazar, U. J. (2017). Linking LFV Higgs decays $ h\to\ell_i\ell_j $ with CP violation in multi-scalar models.\url{https://arxiv.org/abs/1706.00054}

\bibitem{fcnc2}
Alves, J. M., Botella, F. J., Branco, G. C., Cornet-Gomez, F., Nebot, M. (2017). Controlled flavour changing neutral couplings in two Higgs Doublet models. The European Physical Journal C, 77(9), 585.\url{https://doi.org/10.1140/epjc/s10052-017-5156-3}


\bibitem{Martin:1997ns} 
  S.~P.~Martin,
  In *Kane, G.L. (ed.): Perspectives on supersymmetry II* 1-153
  [hep-ph/9709356].

\bibitem{drees}
Drees, M. (1989). Supersymmetric models with extended Higgs sector. International Journal of Modern Physics A, 4(14), 3635-3651.

\bibitem{masip}
Masip, M., Rašin, A. (1995). Spontaneous CP violation in supersymmetric models with four Higgs doublets. Physical Review D, 52(7), R3768.

\bibitem{Aranda:2000zf} 
  A.~Aranda and M.~Sher,
  Phys.\ Rev.\ D {\bf 62}, 092002 (2000)
  doi:10.1103/PhysRevD.62.092002
  [hep-ph/0005113].

 \bibitem{DiazCruz:2002er}
 J.~L.~Diaz-Cruz,
 ``A More flavored Higgs boson in supersymmetric models,''
 JHEP {\bf 0305}, 036 (2003)
 doi:10.1088/1126-6708/2003/05/036
 [hep-ph/0207030].


\bibitem{Dutta:2018yos} 
B.~Dutta and Y.~Mimura,
  Phys.\ Lett.\ B {\bf 790}, 589 (2019)
  doi:10.1016/j.physletb.2019.01.065
  [arXiv:1810.08413 [hep-ph]].
  

\bibitem{kawase} 
Kawase, H. (2011). Light neutralino dark matter scenario in supersymmetric four-Higgs doublet model. Journal of High Energy Physics, 2011(12), 94.
\url{ https://doi.org/10.1007/JHEP12(2011)094}

\bibitem{kanemura} 
S. Kanemura, T. Ota and K. Tsumura, Phys. Rev. D 73, 016006 (2006) \url{http://arxiv.org/abs/hep-ph/0505191}

\bibitem{Gupta:2009wn} 
  R.~S.~Gupta and J.~D.~Wells,
  Phys.\ Rev.\ D {\bf 81}, 055012 (2010)
  doi:10.1103/PhysRevD.81.055012
  [arXiv:0912.0267 [hep-ph]].
  
 
 
 \bibitem{AguilarSaavedra:2004wm}
 J.~A.~Aguilar-Saavedra,
 Acta Phys.\ Polon.\ B {\bf 35}, 2695 (2004)
 [hep-ph/0409342].
 
  \bibitem{uhf} 
  Giardino, P. P., Kannike, K., Masina, I., Raidal, M.,  Strumia, A. (2014). The universal Higgs fit. Journal of High Energy Physics, 2014(5), 46.\url{http://arxiv.org/abs/1303.3570}
 
 \bibitem{diago} 
 A. W. El Kaffas, W. Khater, O. M. Ogreid and P. Osland, Consistency of the two Higgs doublet model and CP violation in top production at the LHC, Nucl. Phys. B 775 (2007) 45
 \url{https:doi:10.1016/j.nuclphysb.2007.03.041 [hep-ph/0605142]}
 
 
\bibitem{tx1} 
T. P. Cheng and M. Sher, Phys. Rev. D 35, 3484 (1987).\url{https://link.aps.org/doi/10.1103/PhysRevD.35.3484}

\bibitem{tx2} 
Y. -F. Zhou, J. Phys. G 30, 783 (2004) [hep-ph/0307240].\url{https://arxiv.org/abs/hep-ph/0307240}

\bibitem{tx3} 
 J. L. Diaz-Cruz, R. Noriega-Papaqui and A. Rosado, Phys. Rev. D 69, 095002 (2004), [hep-ph/0401194].\url{https://arxiv.org/abs/hep-ph/0401194}

 \bibitem{tx4} 
 J. L. Diaz-Cruz, R. Noriega-Papaqui and A. Rosado, Phys. Rev. D 71, 015014 (2005) [hep-ph/0410391].\url{https://arxiv.org/abs/hep-ph/0410391}
 
 
 \bibitem{tx5} 
Y. -L. Wu and Y. -F. Zhou, Eur. Phys. J. C 36, 89 (2004) [hep-ph/0403252].\url{https://arxiv.org/abs/hep-ph/0403252}


\bibitem{tx6} 
W. -j. Li, Y. -d. Yang and X. -d. Zhang, Phys. Rev. D 73, 073005 (2006) [hep-ph/0511273].\url{https://arxiv.org/abs/hep-ph/0511273}

\bibitem{DCT} 
Diaz-Cruz, J. L., Toscano, J. J. (2000). Lepton flavor violating decays of Higgs bosons beyond the standard model. Physical Review D, 62(11), 116005.\url{https://doi.org/10.1016/j.nuclphysb.2016.03.034}

\bibitem{tx7} 
A. E. Carcamo Hernandez, R. Martinez and J. A. Rodriguez, Eur. Phys. J. C 50, 935 (2007) [hep-ph/0606190].\url{https://arxiv.org/pdf/1410.2481.pdf}
  




\bibitem{tx8} 
D. Atwood, S. Bar-Shalom and A. Soni, Phys. Lett. B 635, 112 (2006) doi:10.1016/j.physletb.2006.02.033 [hep-ph/0502234].\url{https://arxiv.org/pdf/1306.2343}

\bibitem{Arroyo:2013tna} 
M.~A.~Arroyo-Ureña, J.~L.~Diaz-Cruz, E.~Díaz and J.~A.~Orduz-Ducuara,
Chin.\ Phys.\ C {\bf 40}, no. 12, 123103 (2016)
doi:10.1088/1674-1137/40/12/123103
[arXiv:1306.2343 [hep-ph]].


\bibitem{Aoki:2011yy} 
  M.~Aoki, S.~Kanemura, T.~Shindou and K.~Yagyu,
  JHEP {\bf 1111}, 038 (2011)
  doi:10.1007/JHEP11(2011)038
  [arXiv:1108.1356 [hep-ph]].
  
\bibitem{Aad:2019mbh}
G.~Aad \textit{et al.} [ATLAS],
Phys. Rev. D \textbf{101} (2020) no.1, 012002
doi:10.1103/PhysRevD.101.012002
[arXiv:1909.02845 [hep-ex]].

\bibitem{Sirunyan:2018koj}
A.~M.~Sirunyan \textit{et al.} [CMS],
Eur. Phys. J. C \textbf{79} (2019) no.5, 421
doi:10.1140/epjc/s10052-019-6909-y
[arXiv:1809.10733 [hep-ex]].

\bibitem{Arroyo-Urena:2020qup}
M.~A.~Arroyo-Ureña, R.~Gaitán and T.~A.~Valencia-Pérez,
[arXiv:2008.00564 [hep-ph]].

\bibitem{PDG}
P.A. Zyla et al. (Particle Data Group), Prog. Theor. Exp. Phys. 2020, 083C01 (2020).

\bibitem{Papaefstathiou:2017xuv}
A.~Papaefstathiou and G.~Tetlalmatzi-Xolocotzi,
Eur. Phys. J. C \textbf{78} (2018) no.3, 214
doi:10.1140/epjc/s10052-018-5701-8
[arXiv:1712.06332 [hep-ph]].

\bibitem{Babu:2018uik} 
  K.~S.~Babu and S.~Jana,
  JHEP {\bf 1902}, 193 (2019)
  doi:10.1007/JHEP02(2019)193
  [arXiv:1812.11943 [hep-ph]].

\bibitem{Hou:2020ciy} 
  W.~S.~Hou, T.~H.~Hsu and T.~Modak,
  Phys.\ Rev.\ D {\bf 102}, no. 5, 055006 (2020)
  doi:10.1103/PhysRevD.102.055006
  [arXiv:2008.02573 [hep-ph]].
  
\bibitem{Altunkaynak:2015twa} 
  B.~Altunkaynak, W.~S.~Hou, C.~Kao, M.~Kohda and B.~McCoy,
  Phys.\ Lett.\ B {\bf 751}, 135 (2015)
  doi:10.1016/j.physletb.2015.10.024
  [arXiv:1506.00651 [hep-ph]].
  
  
\bibitem{Crivellin:2013wna} 
  A.~Crivellin, A.~Kokulu and C.~Greub,
  Phys.\ Rev.\ D {\bf 87}, no. 9, 094031 (2013)
  doi:10.1103/PhysRevD.87.094031
  [arXiv:1303.5877 [hep-ph]].
  
\bibitem{Altmannshofer:2019ogm} 
  W.~Altmannshofer, B.~Maddock and D.~Tuckler,
  Phys.\ Rev.\ D {\bf 100}, no. 1, 015003 (2019)
  doi:10.1103/PhysRevD.100.015003
  [arXiv:1904.10956 [hep-ph]].

 
\bibitem{Arroyo-Urena:2019qhl}
M.~A.~Arroyo-Ureña, R.~Gaitán-Lozano, E.~A.~Herrera-Chacón, J.~H.~Montes de Oca Y. and T.~A.~Valencia-Pérez,
JHEP \textbf{07} (2019), 041
doi:10.1007/JHEP07(2019)041
[arXiv:1903.02718 [hep-ph]].

\bibitem{Bolanos:2019dso}
A.~Bolaños, R.~Sánchez-Vélez and G.~Tavares-Velasco,
Eur. Phys. J. C \textbf{79} (2019) no.8, 700
doi:10.1140/epjc/s10052-019-7211-8
[arXiv:1907.05877 [hep-ph]].

\bibitem{Apollinari:2017cqg}
G.~Apollinari, O.~Brüning, T.~Nakamoto and L.~Rossi,
CERN Yellow Rep. (2015) no.5, 1-19
doi:10.5170/CERN-2015-005.1
[arXiv:1705.08830 [physics.acc-ph]].

\bibitem{Aaboud:2018pob}
M.~Aaboud \textit{et al.} [ATLAS],
Phys. Rev. D \textbf{98} (2018) no.3, 032002
doi:10.1103/PhysRevD.98.032002
[arXiv:1805.03483 [hep-ex]].

\bibitem{Sirunyan:2017uae}
A.~M.~Sirunyan \textit{et al.} [CMS],
JHEP \textbf{06} (2018), 102
doi:10.1007/JHEP06(2018)102
[arXiv:1712.02399 [hep-ex]].

\bibitem{Alwall:2011uj}
J.~Alwall, M.~Herquet, F.~Maltoni, O.~Mattelaer and T.~Stelzer,
JHEP \textbf{06} (2011), 128
doi:10.1007/JHEP06(2011)128
[arXiv:1106.0522 [hep-ph]].

\bibitem{Belyaev:2012qa}
A.~Belyaev, N.~D.~Christensen and A.~Pukhov,
Comput. Phys. Commun. \textbf{184} (2013), 1729-1769
doi:10.1016/j.cpc.2013.01.014
[arXiv:1207.6082 [hep-ph]].

\bibitem{Sjostrand:2006za}
T.~Sjostrand, S.~Mrenna and P.~Z.~Skands,
JHEP \textbf{05} (2006), 026
doi:10.1088/1126-6708/2006/05/026
[arXiv:hep-ph/0603175 [hep-ph]].

\bibitem{deFavereau:2013fsa}
J.~de Favereau \textit{et al.} [DELPHES 3],
JHEP \textbf{02} (2014), 057
doi:10.1007/JHEP02(2014)057
[arXiv:1307.6346 [hep-ex]].

\bibitem{Gao:2013xoa}
J.~Gao, M.~Guzzi, J.~Huston, H.~L.~Lai, Z.~Li, P.~Nadolsky, J.~Pumplin, D.~Stump and C.~P.~Yuan,
Phys. Rev. D \textbf{89} (2014) no.3, 033009
doi:10.1103/PhysRevD.89.033009
[arXiv:1302.6246 [hep-ph]].


\bibitem{Eilam:1990zc} 
G.~Eilam, J.~L.~Hewett and A.~Soni,
Phys.\ Rev.\ D {\bf 44}, 1473 (1991)
Erratum: [Phys.\ Rev.\ D {\bf 59}, 039901 (1999)].
doi:10.1103/PhysRevD.44.1473, 10.1103/PhysRevD.59.039901


\bibitem{DiazCruz:1989ub} 
J.~L.~Diaz-Cruz, R.~Martinez, M.~A.~Perez and A.~Rosado,
Phys.\ Rev.\ D {\bf 41}, 891 (1990).
doi:10.1103/PhysRevD.41.891

\bibitem{DiazCruz:2001gf} 
J.~L.~Diaz-Cruz, H.~J.~He and C.~P.~Yuan,
Phys.\ Lett.\ B {\bf 530}, 179 (2002)
doi:10.1016/S0370-2693(02)01330-8
[hep-ph/0103178].
\bibitem{Sushi1}
SusHi: A program for the calculation of Higgs production in gluon fusion and
bottom-quark annihilation in the Standard Model and the MSSM 
Robert V. Harlander, Stefan Liebler, Hendrik Mantler
Comp. Phys. Commun. 184 (2013) 1605-1617 [arXiv:1212.3249]
DOI: 10.1016/j.cpc.2013.02.006 
\bibitem{Sushi2} 
SusHi Bento: Beyond NNLO and the heavy top-limit 
Robert V. Harlander, Stefan Liebler, Hendrik Mantler
Comp. Phys. Commun. 212 (2017) 239-257 [arXiv:1605.03190]
DOI: 10.1016/j.cpc.2016.10.015 
%
%

 \bibitem{Arroyo-Urena:2019fyd} 
 M.~A.~Arroyo-Ureña, A.~Fernández-Téllez and G.~Tavares-Velasco,
 arXiv:1906.07821 [hep-ph]
 %
 %
 \bibitem{Alwall:2011uj}
 Alwall J, Herquet M, Maltoni F, Mattelaer O, Stelzer T (2011) {MadGraph 5 :
   Going Beyond} 06:128, {10.1007/JHEP06(2011)128}, {1106.0522}.
 
 \bibitem{Alwall:2014hca}
 Alwall J, Frederix R, Frixione S, Hirschi V, Maltoni F, Mattelaer O, Shao HS,
   Stelzer T, Torrielli P, Zaro M (2014) {The automated computation of
   tree-level and next-to-leading order differential cross sections, and their
   matching to parton shower simulations}. JHEP 07:079,
    {10.1007/JHEP07(2014)079}, {1405.0301}.
 
 \bibitem{Sjostrand:2006za}
 Sjostrand T, Mrenna S, Skands PZ (2006) {PYTHIA 6.4 Physics and Manual} 05:026, {10.1088/1126-6708/2006/05/026}, {hep-ph/0603175}.


\end{thebibliography}
\end{document}